\documentclass[aps,prd,twocolumn,superscriptaddress,nofootinbib,floatfix]{revtex4-2}
\usepackage{amsmath,amssymb,mathtools,bm}
\usepackage{booktabs}
\usepackage{graphicx}
\usepackage[colorlinks=true,linkcolor=blue,citecolor=blue,urlcolor=blue]{hyperref}
\usepackage{newunicodechar}
\newunicodechar{ζ}{\ensuremath{\zeta}}

\newcommand{\Pns}{P_{\rm ns}}
\newcommand{\cPns}{\mathcal P_{\rm ns}}
\newcommand{\Amat}{\boldsymbol{\mathcal A}}
\newcommand{\Mmat}{\mathbf M}
\newcommand{\KLG}{extended IREE}
\newcommand{\LL}{{\rm LL}}
\newcommand{\notebelow}[2]{\hphantom{#1}\llap{#2}}

\begin{document}
\title{Non-singlet splitting function at small $x$:\\ a missing two-gluon state in the infrared evolution equations}
\author{Josuke Minamiguchi}
\affiliation{Department of Physics, The University of Tokyo, 7-3-1 Hongo, Bunkyo-ku, Tokyo 113-0033, Japan}

\begin{abstract}
The infrared evolution equations (IREE) resum the double logarithms at small $x$ for general numbers of colors and quark flavors. The conventional IREE give no leading logarithms (LL) for the part $\Pns^{\rm sea}$ of the splitting function for the total valence distribution, the sum over flavors of the differences between quark and antiquark distributions. The three- and four-loop splitting functions, however, contain LL terms in $\Pns^{\rm sea}$. The IREE extended by the symmetric color-octet two-gluon state reproduce these terms with no adjustable parameter and predict the LL at five and six loops. We solve the extended IREE at all orders in closed form at fixed coupling. The solution gives a faster growth of the total valence distribution at small $x$ and larger LL of $F_3$ than the conventional IREE, and it gives a nonzero $s-\bar s$. We consider that the extended IREE are the most promising approach to $\Pns^{\rm sea}$ at small $x$.
\end{abstract}

\maketitle

\section{Introduction}
\label{sec:intro}

At high energies, cross sections in QCD depend on the parton distributions at small momentum fractions $x$. The Electron-Ion Collider (EIC) is designed to measure the parton distributions at small $x$~\cite{Accardi:2012qut,AbdulKhalek:2021gbh}, and predictions for its measurements at small $x$ have been given~\cite{Adamiak:2021ppq,Adamiak:2023yhz}. Below the range of $x$ of the existing data, the parton distributions~\cite{NNPDF:2021njg} and the measurements of several sum rules~\cite{NewMuon:1993oys,Kim:1998kia,COMPASS:2016jwv} rely on extrapolations in $x$.

On the theoretical side, the DGLAP equation~\cite{Gribov:1972ri,Lipatov:1974qm,Altarelli:1977zs,Dokshitzer:1977sg} evolves the parton distributions in the factorization scale with the splitting functions. At small $x$, the highest power of $\ln x$ in the splitting functions increases with the order of the strong coupling. Small-$x$ resummations~\cite{Gelis:2010nm,Kovchegov:2012mbw,Morreale:2021pnn,Hentschinski:2022xnd} sum to all orders the leading logarithms (LL). At each power of the coupling, the LL carry the highest power of $\ln x$ at the most singular power of $x$. For the distributions that mix with the gluon distribution in the evolution, the LL are proportional to $x^{-1}$ and come from gluon exchange in the $t$ channel. The solution of the BFKL equation~\cite{Fadin:1975cb,Lipatov:1976zz,Kuraev:1977fs,Balitsky:1978ic} gives these LL~\cite{Jaroszewicz:1982gr}. In contrast, the sum over flavors of the difference between the quark and antiquark distributions does not mix with the gluon distribution. This sum is called the total valence distribution~\cite{Moch:2004pa}. The splitting function for the total valence distribution has no $x^{-1}$ terms~\cite{Catani:1994sq}, and its LL are $x^0$ terms with two powers of $\ln x$ for each power of the coupling~\cite{Kirschner:1983di,Blumlein:1995jp}. Terms of this form are called double logarithms (DL). The LL of the splitting functions for the helicity distributions are also DL~\cite{Bartels:1996wc}. The BFKL equation and its nonlinear extensions, the BK~\cite{Balitsky:1995ub,Kovchegov:1999yj} and JIMWLK~\cite{Jalilian-Marian:1997jhx,Iancu:2000hn,Weigert:2000gi} equations, resum $x^{-1}$ terms and do not resum the DL~\cite{Kovchegov:2016zex}.

The infrared evolution equations (IREE)~\cite{Kirschner:1982qf,Kirschner:1983di,Ermolaev:2007yb,Lipatov:2017inx,Ermolaev:2026rkl} resum the DL for general numbers of colors and quark flavors~\cite{Kirschner:1983di,Bartels:1996wc,Blumlein:1996dd}. For example, the IREE give the small-$x$ behavior of the part of $F_1$ that depends only on the differences between the quark distributions of different flavors~\cite{Ermolaev:1995fx} and of the polarized structure function $g_1$~\cite{Bartels:1995iu,Bartels:1996wc}. For the differences between the distributions of two quark flavors, the LL of the IREE agree with the three-loop~\cite{Moch:2004pa,Blumlein:2021enk} and four-loop~\cite{Moch:2026qsw} splitting functions.

On the other hand, in the splitting function for the total valence distribution, the three-loop~\cite{Moch:2004pa} and four-loop~\cite{Gehrmann:2026qbl} calculations found LL terms proportional to the color factor $d^{abc}d_{abc}$, where $d^{abc}$ is the totally symmetric invariant tensor of the color group. The IREE had not predicted the $d^{abc}d_{abc}$ terms~\cite{Moch:2004pa,Davies:2016jie}. The $d^{abc}d_{abc}$ terms belong to the sea part $\Pns^{\rm sea}$ [Eq.~\eqref{eq:Ps}] of the splitting function for the total valence distribution. To our knowledge, no method in the literature gives the LL of $\Pns^{\rm sea}$ at all orders. We therefore consider it important to obtain the all-order LL of $\Pns^{\rm sea}$ before the EIC starts operation. Other problems in the resummation of the DL are discussed in Refs.~\cite{Cougoulic:2022gbk,Borden:2023ugd,Borden:2024bxa,Borden:2025ehe,Hatta:2026txn}.

In this paper, we extend the IREE so that they resum the LL of $\Pns^{\rm sea}$ to all orders at fixed coupling. We call the IREE of Refs.~\cite{Kirschner:1983di,Bartels:1995iu,Blumlein:1996dd} the conventional IREE. In the conventional IREE, the LL of $\Pns^{\rm sea}$ vanish at every power of the coupling, because the $t$-channel intermediate pairs of the conventional IREE (Sec.~\ref{sec:iree}) are quark--antiquark pairs only. We add to the conventional IREE a symmetric color-octet state of two gluons, whose colors are coupled by $d^{abc}$, and we call the resulting equations the \KLG{}. With no adjustable parameter, the \KLG{} reproduce the three- and four-loop LL of $\Pns^{\rm sea}$ for any number of colors.

At fixed coupling, we also use the \KLG{} to obtain the all-order LL of the strange-quark asymmetry $s-\bar s$ and of the structure function $F_3$ summed over neutrino and antineutrino charged-current scattering. The evolution with $\Pns^{\rm sea}$ generates $s-\bar s$ even if $s=\bar s$ at the initial scale, and this asymmetry has been calculated with the three-loop $\Pns^{\rm sea}$~\cite{Catani:2004nc}. The value of $\sin^2\theta_W$ for the weak mixing angle $\theta_W$ obtained from neutrino scattering~\cite{NuTeV:2001whx} depends on $\int_0^1dx\,x(s-\bar s)$~\cite{Davidson:2001ji,Kretzer:2003wy}, and experimental data constrain $s$ and $\bar s$ separately only at $x\gtrsim0.01$~\cite{NuTeV:2001dfo,Ball:2018twp,Anderson:2024evk}. At leading order, $x$ times this $F_3$ is proportional to $x$ times the total valence distribution~\cite{CCFR:1997tam}. The data on $xF_3$ start at $x=0.0075$~\cite{CCFR:1997tam}, and the measurement~\cite{Kim:1998kia} of the sum rule for the integral of $F_3$ over $x$~\cite{Gross:1969jf} extrapolates $F_3$ below this value of $x$. We give the all-order LL of this $F_3$ for a quark target in the factorization scheme defined by the infrared cutoff of the IREE on the transverse momenta of virtual partons (Sec.~\ref{sec:f3}).

Section~\ref{sec:setup} defines $\Pns^{\rm sea}$ and reviews it at fixed order and in the conventional IREE. Section~\ref{sec:eq} adds the symmetric color-octet two-gluon state and derives the \KLG{}. Section~\ref{sec:closed} solves the \KLG{} at all orders and gives the small-$x$ growth of the total valence distribution, $\Pns^{\rm sea}$ in $x$ space, the strange-quark asymmetry and $F_3$. Section~\ref{sec:fixed} compares the solution with the three- and four-loop results and gives predictions at five and six loops. Section~\ref{sec:discussion} discusses the results.

\section{The splitting function \texorpdfstring{$\Pns^{\rm sea}$}{P\_ns\textasciicircum sea} at fixed order and in the conventional IREE}
\label{sec:setup}

\subsection{Valence and sea parts of the splitting functions}
\label{sec:valsea}
We write $q_k$ and $\bar q_k$ for the quark and the antiquark of flavor $k$ and for their distributions, and $n_f$ for the number of quark flavors. The quark distribution $\sum_k(q_k+\bar q_k)$ and the gluon distribution mix in the evolution and are called the flavor singlet. The combinations of quark and antiquark distributions that do not mix with the gluon distribution are called non-singlet. We expand every splitting function $P$ in powers of $a_s=\alpha_s/(4\pi)$, where $\alpha_s$ is the strong coupling:
\begin{equation}
P=\sum_{n\ge0}a_s^{n+1}\,P^{(n)}.
\label{eq:conv}
\end{equation}
We write the Mellin transform of every splitting function $P$ as
\begin{equation}
\mathcal P(N)=\int_0^1dx\,x^{N-1}P(x).
\label{eq:mellin}
\end{equation}
The Mellin transform $\mathcal P(N)$ of a splitting function is called an anomalous dimension.

We write the splitting functions from a parent quark $q_k$ and from a parent antiquark $\bar q_k$ to a daughter quark $q_i$ as $P_{q_iq_k}$ and $P_{q_i\bar q_k}$, respectively. Flavor symmetry and charge conjugation give the decompositions~\cite{Moch:2004pa}
\begin{align}
&P_{q_iq_k}=\delta_{ik}P_{qq}^{\rm v}+P_{qq}^{\rm s},\label{eq:flavor}\\
&P_{q_i\bar q_k}=\delta_{ik}P_{q\bar q}^{\rm v}+P_{q\bar q}^{\rm s}.\label{eq:flavorbar}
\end{align}
For any matrix in flavor space, the part proportional to $\delta_{ik}$ and the part that takes the same value for all $i$ and $k$ are called the valence part and the sea part, respectively~\cite{Moch:2004pa}. In Eqs.~\eqref{eq:flavor} and \eqref{eq:flavorbar}, $P_{qq}^{\rm v}$ and $P_{q\bar q}^{\rm v}$ are the valence parts, and $P_{qq}^{\rm s}$ and $P_{q\bar q}^{\rm s}$ are the sea parts. The total valence distribution $\sum_k(q_k-\bar q_k)$ evolves with the splitting function~\cite{Moch:2004pa}
\begin{align}
&\Pns^{\rm val}=\Pns^{-}+\Pns^{\rm sea},\label{eq:Pv}\\
&\Pns^{-}=P_{qq}^{\rm v}-P_{q\bar q}^{\rm v},\label{eq:Pminus}\\
&\Pns^{\rm sea}=n_f\left(P_{qq}^{\rm s}-P_{q\bar q}^{\rm s}\right).\label{eq:Ps}
\end{align}
The difference $P_{q_iq_k}-P_{q_i\bar q_k}$ between the contributions of a parent quark $q_k$ and of a parent antiquark $\bar q_k$ has the valence part $\Pns^{-}$ and the sea part $\Pns^{\rm sea}/n_f$. We call such a difference the parent combination $q-\bar q$ and the corresponding sum the parent combination $q+\bar q$. The valence part of the sum is $\Pns^{+}=P_{qq}^{\rm v}+P_{q\bar q}^{\rm v}$. The splitting functions $\Pns^+$ and $\Pns^-$ evolve the differences $(q_i+\bar q_i)-(q_k+\bar q_k)$ and $(q_i-\bar q_i)-(q_k-\bar q_k)$ between flavors, respectively. The superscripts val and sea label quantities summed over the daughter flavor $i$. The sum of $P_{q_iq_k}-P_{q_i\bar q_k}$ over $i$ is $\Pns^{\rm val}$, and the sum of the sea part of this difference over $i$ is $\Pns^{\rm sea}$.

\begin{figure*}[t]
\centering
\includegraphics[page=1]{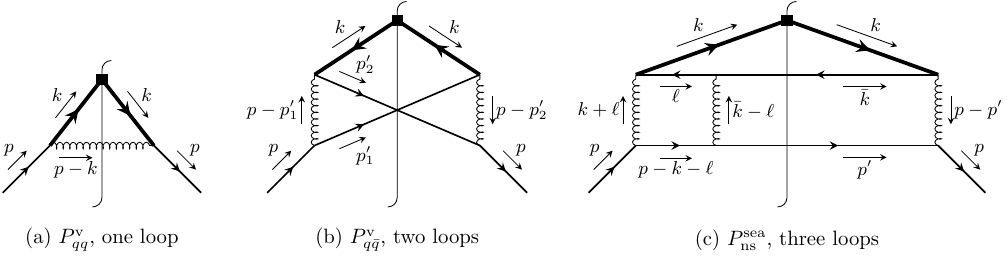}
\caption{Cut diagrams of $P_{qq}^{\rm v}$ at $a_s$ (a), $P_{q\bar q}^{\rm v}$ at $a_s^2$ (b) and $\Pns^{\rm sea}$ at $a_s^3$ (c). The black square is the vertex of the operator that defines the parton distribution, and thick lines are the observed parton with the momentum $k$.}
\label{fig:valsea}
\end{figure*}

Figure~\ref{fig:valsea} shows a cut diagram of $P_{qq}^{\rm v}$ at $a_s$, of $P_{q\bar q}^{\rm v}$ at $a_s^2$ and of $\Pns^{\rm sea}$ at $a_s^3$. These powers are the lowest powers of $a_s$ in the three splitting functions. In the diagrams of this paper, the $t$ channel runs from bottom to top, and time runs from left to right. In Fig.~\ref{fig:valsea}, the black square is the vertex $\gamma^+\delta(k^+-xp^+)/2$ of the operator that defines the parton distribution~\cite{Collins:1981uw}, where $k^+$ and $p^+$ are the plus momenta of the observed parton and of the parent. In a valence part, the observed parton lies on the fermion line of the parent [Figs.~\ref{fig:valsea}(a), (b)]. In a sea part, the observed parton lies on a closed fermion loop attached to the parent line by gluons [Fig.~\ref{fig:valsea}(c)].

\subsection{Leading small-\texorpdfstring{$x$}{x} terms of \texorpdfstring{$\Pns^{\rm sea}$}{P\_ns\textasciicircum sea} at three and four loops}
\label{sec:data}
The three-loop $\Pns^{\rm sea}$ is proportional to $d^{abc}d_{abc}$ of SU($N_c$)~\cite{Moch:2004pa}, where $N_c$ is the number of colors. By charge conjugation, $P_{q\bar q}^{\rm s}$ equals the sea part of the splitting function from a parent quark to a daughter antiquark. The daughter antiquark lies on a loop with the fermion flow opposite to that of the loop in $P_{qq}^{\rm s}$. Reversing the fermion flow of a loop reverses the order of the generators $T^a$ of SU($N_c$) in the fundamental representation and gives a factor $-1$ for each gluon vertex. In Fig.~\ref{fig:valsea}(c), the loop of the observed quark $q_i$ and the loop with the opposite fermion flow have, respectively, the color factors
\begin{align}
&{\rm tr}(T^aT^bT^c)=\frac14\left(d^{abc}+if^{abc}\right),\label{eq:tr3}\\
&(-1)^3\,{\rm tr}(T^cT^bT^a)=-\frac14\left(d^{abc}-if^{abc}\right).\label{eq:tr3rev}
\end{align}
Here $f^{abc}$ are the structure constants of SU($N_c$). In the difference $P_{qq}^{\rm s}-P_{q\bar q}^{\rm s}$, only the $d^{abc}$ terms remain, and their contraction with the color factor of the parent line is proportional to $d^{abc}d_{abc}$.

In the non-singlet sector, the LL (Sec.~\ref{sec:intro}) are the DL $a_s^n\,x^0\ln^{2n-2}x$~\cite{Kirschner:1983di,Bartels:1995iu,Blumlein:1995jp}.

The fixed-order calculations of $\Pns^{\rm sea}$ at three~\cite{Moch:2004pa} and four~\cite{Gehrmann:2026qbl} loops give the LL
\begin{align}
&\Pns^{(2){\rm sea}}(x)\Big|_{\ln^4x}=\frac13\,n_fD\,\ln^4x\qquad(\text{fixed order}),\label{eq:data3}\\
&\Pns^{(3){\rm sea}}(x)\Big|_{\ln^6x}=n_fD\left(\frac{2}{15}C_F-\frac15C_A\right)\ln^6x\nonumber\\
&\notebelow{\Pns^{(3){\rm sea}}(x)\Big|_{\ln^6x}=n_fD\left(\frac{2}{15}C_F-\frac15C_A\right)\ln^6x}{(fixed order),}\label{eq:data4}
\end{align}
where $C_F=(N_c^2-1)/(2N_c)$, $C_A=N_c$ and $D$ is the color factor
\begin{equation}
D=\frac{d^{abc}d_{abc}}{N_c}=\frac{2C_F(N_c^2-4)}{N_c}.
\label{eq:D}
\end{equation}

The correspondence $\ln^{2n-2}x\leftrightarrow(2n-2)!/N^{2n-1}$ between $x$ space and Mellin space turns Eqs.~\eqref{eq:data3} and \eqref{eq:data4} into
\begin{equation}
\begin{split}
\cPns^{\rm sea}(N)\Big|_\LL=n_fD\biggl[&\frac{8\,a_s^3}{N^5}+\left(96C_F-144C_A\right)\frac{a_s^4}{N^7}\\
&+O\!\left(\frac{a_s^5}{N^9}\right)\biggr]\qquad(\text{fixed order}).
\end{split}
\label{eq:dataN}
\end{equation}
Here $O(a_s^5/N^9)$ stands for the LL terms from five loops on. These terms have not been calculated at fixed order.

\subsection{Splitting functions and the partial waves of \texorpdfstring{$qq\to qq$}{qq->qq}}
\label{sec:pw}
\begin{figure}[t]
\centering
\includegraphics[page=2]{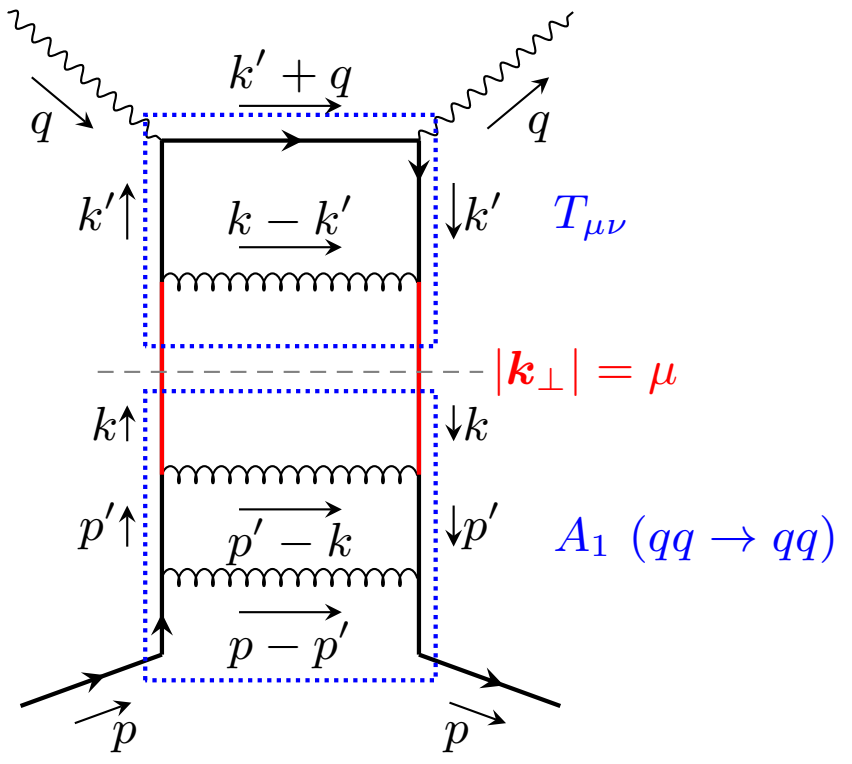}
\caption{Division of the forward amplitude $T_{\mu\nu}$ of deep-inelastic scattering on a quark target by the softest $t$-channel intermediate pair (red lines). For the component $T_3$ [Eq.~\eqref{eq:Tdecomp}], the upper part is $T_3$ for a target quark of the flavor of the pair, and the lower part is $A_1^-$.}
\label{fig:readout}
\end{figure}

The diagrams of Fig.~\ref{fig:valsea} with two currents in place of the black square are cuts of the forward amplitude of deep-inelastic scattering on a quark target (Fig.~\ref{fig:readout}). This amplitude is
\begin{equation}
T_{\mu\nu}=i\int d^4z\,e^{iq\cdot z}\,\langle p|{\rm T}\{J_\mu^\dagger(z)J_\nu(0)\}|p\rangle,
\label{eq:Tmunu}
\end{equation}
where $J_\mu$ is the electromagnetic or charged weak current, $q$ and $p$ are the momenta of the current and of the target quark, $Q^2=-q^2$, $x=Q^2/(2p\cdot q)$, and we average over the spin of the target quark. For $x\ll1$, we approximate the squared center-of-mass energy $s=(p+q)^2$ by $Q^2/x$. At leading power in $1/Q^2$, the cut quark line between the two currents is proportional to $\gamma^+\delta(k^+-xp^+)$~\cite{Collins:2011zzd}. We impose a lower cutoff $\mu$ on the transverse momenta of all virtual partons.

We consider the structure function $F_3$ of charged-current deep-inelastic scattering summed over neutrino and antineutrino scattering. We compute $F_3(x,Q^2/\mu^2)$ of a quark target with the cutoff $\mu$ at fixed coupling. We write the forward amplitude~\eqref{eq:Tmunu} as
\begin{equation}
T_{\mu\nu}=-\frac{i\epsilon_{\mu\nu\alpha\beta}q^\alpha p^\beta}{p\cdot q}\,T_3(\rho,Q^2/\mu^2)+\cdots,
\label{eq:Tdecomp}
\end{equation}
where $\rho=\ln(s/\mu^2)$ and the dots stand for the other tensor structures. The normalization of the tensor structure in Eq.~\eqref{eq:Tdecomp} is that of Ref.~\cite{Moch:2008fj}. For $s>0$, the optical theorem relates $T_3$ to $F_3$ as
\begin{equation}
F_3(x,Q^2/\mu^2)=\frac{1}{2\pi}\,{\rm Im}\,T_3(\rho,Q^2/\mu^2).
\label{eq:F3T3}
\end{equation}

The parts of an amplitude that are symmetric and antisymmetric under the crossing $s\leftrightarrow u$ are called the parts with signature $+$ and $-$, respectively. The crossing $s\leftrightarrow u$ turns a parent quark into a parent antiquark. For a color-singlet $t$ channel, the parent combinations $q\pm\bar q$ therefore have signature $\pm$. With the signature $-$ of $q-\bar q$, we write $T_3$ as
\begin{equation}
T_3(\rho,Q^2/\mu^2)=\int\frac{d\omega}{2\pi i}\,e^{\omega\rho}\,\xi^{(-)}(\omega)\,R(\omega,Q^2/\mu^2),
\label{eq:TR}
\end{equation}
where $R$ is the partial wave of $T_3$ and $\xi^{(\pm)}$ are the signature factors
\begin{equation}
\xi^{(\pm)}(\omega)=\frac{e^{-i\pi\omega}\pm1}{2}.
\label{eq:xi}
\end{equation}

At LL, the transverse momenta of the $n$ virtual partons of a diagram are strongly ordered~\cite{Kirschner:1983di}:
\begin{equation}
\mu^2<\boldsymbol k_{\perp\sigma(1)}^2\ll\boldsymbol k_{\perp\sigma(2)}^2\ll\cdots\ll\boldsymbol k_{\perp\sigma(n)}^2
\label{eq:order}
\end{equation}
for an ordering $\sigma$ of the partons. The ordering $\sigma$ need not follow the positions of the partons along the ladder. A region in which two transverse momenta are comparable gives one logarithm fewer.

At one loop, the DL of a forward amplitude with the squared center-of-mass energy $s$ comes from the exchanged quark with the transverse momentum $\boldsymbol k_\perp$ and the fraction $\beta$ of the longitudinal momentum of one incoming parton. The cutoff gives the lower limit $\boldsymbol k_\perp^2>\mu^2$. The upper limit $\boldsymbol k_\perp^2<\beta s$ is the condition that the squared invariant mass of the exchanged quark and the other incoming parton is non-negative. Up to the color factor, the DL is therefore
\begin{equation}
a_s\int_{\mu^2/s}^{1}\frac{d\beta}{\beta}\int_{\mu^2}^{\beta s}\frac{d\boldsymbol k_\perp^2}{\boldsymbol k_\perp^2}=\frac{a_s}{2}\ln^2\frac{s}{\mu^2}.
\label{eq:DLint}
\end{equation}
At higher powers of $a_s$, each rung gives an integral of the form of Eq.~\eqref{eq:DLint}.

The parton $\sigma(1)$ in Eq.~\eqref{eq:order} is called the softest parton. In Eq.~\eqref{eq:order}, the cutoff $\mu$ bounds only the transverse momentum of the softest parton. The IREE~\cite{Kirschner:1982qf,Kirschner:1983di,Bartels:1995iu} therefore obtain the LL of the derivative $-\partial/\partial\ln\mu^2$ of an amplitude from the softest parton with $|\boldsymbol k_\perp|=\mu$. We call a pair of lines crossed by a horizontal cut of a ladder diagram a $t$-channel intermediate pair. Soft gluons attached to the quark lines next to the currents give no power of $\ln\mu^2$, because these lines are off shell by $Q^2$. The softest parton in Fig.~\ref{fig:readout} therefore belongs to a $t$-channel intermediate pair~\cite{Bartels:1995iu}. The softest $t$-channel intermediate pair divides the amplitude into two parts. Because the currents are color singlets, the softest $t$-channel intermediate pair is a color singlet.

The lower of the two parts is a forward $qq\to qq$ amplitude (Fig.~\ref{fig:readout}). For $qq\to qq$, we write the part whose $t$ channel is a color singlet as $A_1$ and the part whose $t$ channel is a color octet as $A_8$. We write the parts of $A_{1,8}$ with signature $\tau=\pm$ as $A_{1,8}^{\tau}$. For a color-octet $t$ channel, the color tensor contains one more generator $T^a$, and the reversal of the fermion flow of the parent line under the crossing gives one more factor $-1$ (Sec.~\ref{sec:data}). For a color-octet $t$ channel, the parent combinations $q\pm\bar q$ therefore have signature $\mp$. In the same form as Eq.~\eqref{eq:TR}, we write $A_{1,8}^{\tau}$ with the partial waves $\mathcal A_{1,8}^{\tau}$ as
\begin{equation}
A_{1,8}^{\tau}(\rho)=\int\frac{d\omega}{2\pi i}\,e^{\omega\rho}\,\xi^{(\tau)}(\omega)\,\mathcal A_{1,8}^{\tau}(\omega).
\label{eq:pw}
\end{equation}
As in Eq.~\eqref{eq:conv}, we expand the partial waves of $qq\to qq$ as
\begin{equation}
\mathcal A_{1,8}^{\tau}(\omega)=\sum_{n\ge0}a_s^{n+1}\,\mathcal A_{1,8}^{(n)\tau}(\omega).
\label{eq:pwexp}
\end{equation}

The softest $t$-channel intermediate pair divides $T_3$ into the upper part and the lower part $A_1^-$ with the pair as on-shell external lines (Fig.~\ref{fig:readout}). The upper part is $T_3$ for a target quark of the flavor of the pair. Equations~\eqref{eq:pw} and \eqref{eq:TR} separate from the partial waves the phases of the energy logarithms. The signature factor $\xi^{(-)}(\omega)$ in Eq.~\eqref{eq:TR} fixes these phases for the whole of $T_3$. The pair contributes the product of the partial waves of the two parts, without signature factors~\cite{Kirschner:1983di,Bartels:1995iu}. The LL of the derivative $-\partial/\partial\ln\mu^2$ of $T_3$ at fixed $s$ and $Q^2$ are therefore
\begin{equation}
\begin{split}
-\frac{\partial T_3(\rho,Q^2/\mu^2)}{\partial\ln\mu^2}={}&\int\frac{d\omega}{2\pi i}\,e^{\omega\rho}\,\xi^{(-)}(\omega)\\
&\times R(\omega,Q^2/\mu^2)\,\mathcal A_1^-(\omega).
\end{split}
\label{eq:ireeT}
\end{equation}
On the right-hand side of Eq.~\eqref{eq:TR}, the derivative $-\partial/\partial\ln\mu^2$ at fixed $s$ and $Q^2$ multiplies $e^{\omega\rho}$ by $\omega$ and acts on $R$ as $\partial/\partial\ln(Q^2/\mu^2)$. Equation~\eqref{eq:ireeT} therefore gives the infrared evolution equation for the partial wave $R$, which has the same form as that for $g_1$ in Ref.~\cite{Bartels:1995iu}:
\begin{equation}
\begin{split}
&\omega R(\omega,Q^2/\mu^2)+\frac{\partial R(\omega,Q^2/\mu^2)}{\partial\ln(Q^2/\mu^2)}\\
&\quad=R(\omega,Q^2/\mu^2)\,\mathcal A_1^-(\omega).
\end{split}
\label{eq:ireeR}
\end{equation}
We now relate $\mathcal A_1^-$ to the anomalous dimension $\cPns^{-}$. We write the Mellin moments of $F_3$ as
\begin{equation}
\mathcal F_3(N,Q^2/\mu^2)=\int_0^1dx\,x^{N-1}F_3(x,Q^2/\mu^2).
\label{eq:FN}
\end{equation}
In Eq.~\eqref{eq:F3T3}, the factor $\xi^{(-)}(\omega)$ of Eq.~\eqref{eq:TR} gives ${\rm Im}\,\xi^{(-)}(\omega)=-\sin(\pi\omega)/2$. With $s\simeq Q^2/x$ for $x\ll1$, the factor $e^{\omega\rho}$ becomes $(Q^2/\mu^2)^\omega x^{-\omega}$, and $F_3$ becomes an integral over $\omega$ of $x^{-\omega}$ times a function of $\omega$. The Mellin transform~\eqref{eq:mellin} turns $x^{-\omega}$ into $1/(N-\omega)$, and closing the contour on this pole sets $\omega=N$. The moments are therefore
\begin{equation}
\mathcal F_3(N,Q^2/\mu^2)=-\frac{\sin(\pi N)}{4\pi}\left(\frac{Q^2}{\mu^2}\right)^{N}R(N,Q^2/\mu^2).
\label{eq:FR}
\end{equation}
Equations~\eqref{eq:FR} and \eqref{eq:ireeR} give
\begin{equation}
-\frac{\partial\mathcal F_3(N,Q^2/\mu^2)}{\partial\ln\mu^2}=\mathcal A_1^-(N)\,\mathcal F_3(N,Q^2/\mu^2).
\label{eq:ireeF}
\end{equation}
The left-hand side of Eq.~\eqref{eq:ireeF} equals $\partial\mathcal F_3(N,Q^2/\mu^2)/\partial\ln Q^2$. By the DGLAP equation, this derivative is $\cPns^{-}(N)\,\mathcal F_3(N,Q^2/\mu^2)$. Equation~\eqref{eq:ireeF} then gives
\begin{equation}
\cPns^{-}(N)\Big|_\LL=\mathcal A_1^-(N).
\label{eq:readout}
\end{equation}
In the same way, the LL of $\cPns^{+}$ are the partial wave $\mathcal A_1^{+}$ at $\omega=N$~\cite{Bartels:1995iu,Blumlein:1996dd}.

For the parent combination $q-\bar q$, the color-singlet partial waves with a parent of flavor $k$ and a daughter of flavor $i$ form an $n_f\times n_f$ matrix $\Amat_1^-(\omega)$ in flavor space. We write its valence part and its sea part as $\delta_{ik}\mathcal A_1^-(\omega)$ and $\mathcal A_1^{\rm sea}(\omega)/n_f$. At $\omega=N$, this matrix gives the LL of the anomalous dimensions for the parent combination $q-\bar q$:
\begin{equation}
\begin{split}
(\Amat_1^-)_{ik}(N)&=\Bigl[\mathcal P_{q_iq_k}(N)-\mathcal P_{q_i\bar q_k}(N)\Bigr]_\LL\\
&=\delta_{ik}\,\mathcal A_1^-(N)+\frac{\mathcal A_1^{\rm sea}(N)}{n_f}.
\end{split}
\label{eq:readout_s}
\end{equation}
Equations~\eqref{eq:flavor}, \eqref{eq:flavorbar} and \eqref{eq:Ps} then give
\begin{equation}
\cPns^{\rm sea}(N)\Big|_\LL=\mathcal A_1^{\rm sea}(N).
\label{eq:readout_s2}
\end{equation}

\subsection{The conventional IREE}
\label{sec:iree}
By Eqs.~\eqref{eq:readout} and \eqref{eq:readout_s2}, the LL of $\cPns^{-}$ and $\cPns^{\rm sea}$ are given by the color-singlet partial waves of $qq\to qq$. The conventional IREE for $qq\to qq$ determine these partial waves~\cite{Kirschner:1983di,Bartels:1995iu}.

\begin{figure*}[t]
\centering
\includegraphics[page=3]{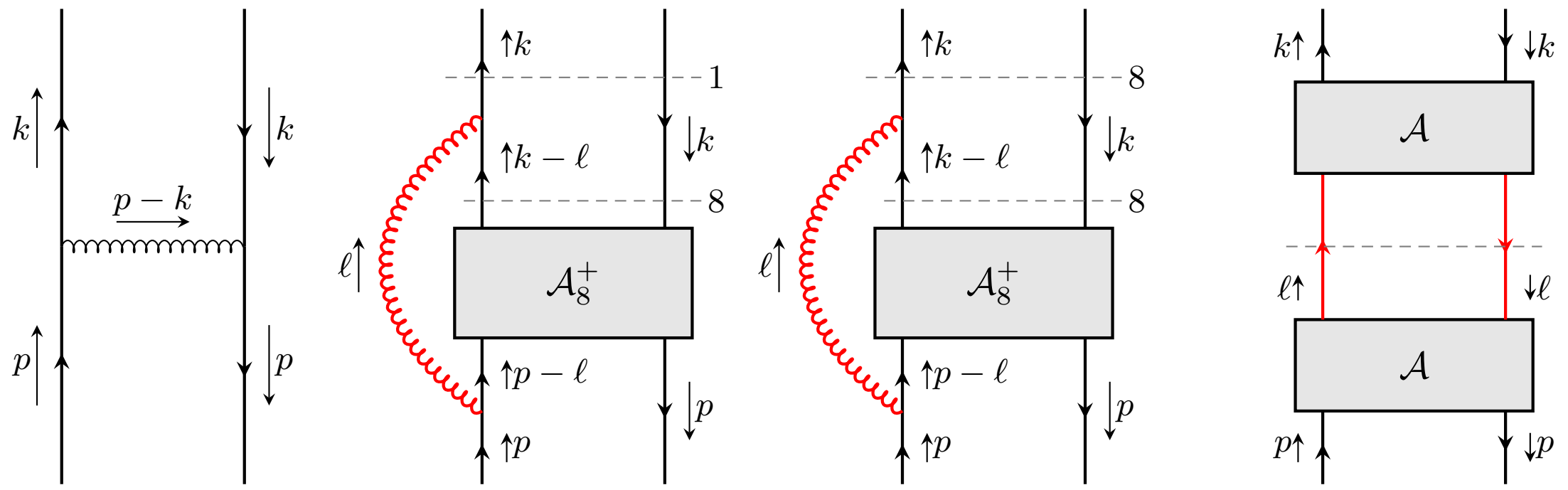}
\caption{Diagrams of the four kinds of terms on the right-hand sides of Eqs.~\eqref{eq:KLp}--\eqref{eq:KL8}. Gray boxes are partial waves, and red lines are the softest partons. The label at each dashed line is the color of the two lines that it crosses.}
\label{fig:terms}
\end{figure*}

For the $qq\to qq$ amplitude, the derivative $-\partial/\partial\ln\mu^2$ at fixed $s$ equals $\partial/\partial\rho$. In Eq.~\eqref{eq:pw}, the derivative $\partial/\partial\rho$ multiplies $e^{\omega\rho}$ by $\omega$. The left-hand side of each of the conventional IREE is therefore $\omega$ times a partial wave. The right-hand side is the sum of the LL of the derivative from three kinds of diagrams (Fig.~\ref{fig:terms}): the diagram with one rung, the diagrams in which the softest gluon is attached to the external quarks, and the diagrams in which the softest $t$-channel intermediate pair divides the ladder into two parts. At LL, the conventional IREE read
\begin{align}
\omega\mathcal A_1^+(\omega)&=2C_Fa_s+[\mathcal A_1^+(\omega)]^2,\label{eq:KLp}\\
\omega\mathcal A_1^-(\omega)&=2C_Fa_s-\frac{8C_Fa_s}{\omega}\,\mathcal A_8^+(\omega)+[\mathcal A_1^-(\omega)]^2,\label{eq:KLm}\\
\omega\mathcal A_8^+(\omega)&=(2C_F-C_A)a_s+2C_Aa_s\,\frac{d\mathcal A_8^+(\omega)}{d\omega}+[\mathcal A_8^+(\omega)]^2.\label{eq:KL8}
\end{align}
The first term on the right-hand side of each equation is the Born term and comes from the diagram with one rung.

The second terms on the right-hand sides of Eqs.~\eqref{eq:KLm} and \eqref{eq:KL8} come from the diagrams in which the softest gluon is attached to the external quarks~\cite{Kirschner:1983di,Bartels:1995iu}. We call them the signature-changing term and the Sudakov term, respectively. The factor $1/\omega$ in the signature-changing term comes from the ratio $\xi^{(+)}/\xi^{(-)}=i\cot(\pi\omega/2)\simeq2i/(\pi\omega)$ for $\omega\ll1$~\cite{Kirschner:1983di}. In the Sudakov term, the derivative $d/d\omega$ corresponds to multiplying $A_8^+(\rho)$ by $-\rho$ [Eq.~\eqref{eq:pw}]. The factor $\rho$ comes from the integral $\int_{\mu^2/s}^{1}d\beta/\beta$ over the longitudinal momentum fraction $\beta$ of the softest gluon~\cite{Kirschner:1983di}.

The last term on the right-hand side of each equation is the ladder term and comes from the diagrams in which the softest $t$-channel intermediate pair divides the ladder into two parts. As in Eq.~\eqref{eq:ireeR}, the pair contributes the product of the partial waves of the two parts.

The conventional IREE give no sea part at any power of $a_s$:
\begin{equation}
\mathcal A_1^{\rm sea}(\omega)=0\qquad(\text{conventional IREE}).
\label{eq:KLzero}
\end{equation}
In Eqs.~\eqref{eq:KLm} and \eqref{eq:KL8} (Fig.~\ref{fig:terms}), every $t$-channel intermediate pair is a $q\bar q$ pair whose two fermion lines continue into external quarks of the same flavor, so that every term is proportional to the unit matrix in flavor space. By Eq.~\eqref{eq:readout_s2}, Eq.~\eqref{eq:KLzero} contradicts Eq.~\eqref{eq:dataN}. The conventional IREE therefore do not resum the LL of $\Pns^{\rm sea}$.

The missing contribution has a symmetric color-octet $t$-channel intermediate pair of two gluons. In Fig.~\ref{fig:valsea}(c), let one of the three gluons be the softest parton. The other two gluons then form a $t$-channel intermediate pair. By Eqs.~\eqref{eq:tr3} and \eqref{eq:tr3rev}, the difference $P_{qq}^{\rm s}-P_{q\bar q}^{\rm s}$ couples the two gluons $a$ and $b$ of the pair to the color $c$ of the softest gluon only through $d^{abc}$. We write $(gg)_{8_s}$ for the color-octet state of two gluons with the colors $a$ and $b$ and the color tensor $d^{abc}$, where $c$ is the color of the octet. The subscript $8_s$ refers to the symmetry of $d^{abc}$ under $a\leftrightarrow b$. The conventional IREE do not contain $(gg)_{8_s}$. The symmetric color-octet state of two reggeized gluons has been studied in single-logarithmic Regge theory~\cite{Kwiecinski:1979wr,Ewerz:2001fb,Kovner:2005qj,Fadin:2005zj,Caron-Huot:2013fea,Caron-Huot:2017zfo,Gao:2024fyz}.

\section{IREE with the symmetric color-octet two-gluon state}
\label{sec:eq}

\subsection{The added state \texorpdfstring{$(gg)_{8_s}$}{(gg)\_8s}}
\label{sec:state}
Below, we show that the state $(gg)_{8_s}$ has positive signature and couples to the parent combination $q-\bar q$. Through the signature-changing term, $(gg)_{8_s}$ therefore contributes at LL to the matrix $\Amat_1^-$ of Eq.~\eqref{eq:readout_s}. The rest of this subsection shows that no other gluon pair contributes at LL to $\Pns^{\pm}$ or $\Pns^{\rm sea}$.

The total color of the $t$ channel is conserved along the ladder. A gluon pair that mixes with a $q\bar q$ pair therefore has the color of the $q\bar q$ pair, a singlet or an octet. Two gluons form the color singlet $(gg)_1$ with $\delta^{ab}$, the symmetric color octet $(gg)_{8_s}$ with $d^{abc}$ and the antisymmetric color octet $(gg)_{8_a}$ with $f^{abc}$. The other representations of two gluons ($10$, $\overline{10}$ and $27$ for $N_c=3$) do not mix with a $q\bar q$ pair.

A color-octet $t$-channel intermediate pair enters the singlet partial waves only through the signature-changing term (Fig.~\ref{fig:terms}), and this term contributes at LL only to $q-\bar q$. The signature-changing term feeds a positive-signature octet into the negative-signature singlet $\mathcal A_1^-$ of $q-\bar q$ with the factor $\xi^{(+)}/\xi^{(-)}=i\cot(\pi\omega/2)$ and gives the LL term $-8C_Fa_s\mathcal A_8^+(\omega)/\omega$ of Eq.~\eqref{eq:KLm}. The signature-changing term also feeds a negative-signature octet into the positive-signature singlet $\mathcal A_1^+$ of $q+\bar q$ with the factor $\xi^{(-)}/\xi^{(+)}=-i\tan(\pi\omega/2)$. For $\omega\to0$, the factor $\xi^{(-)}/\xi^{(+)}$ is smaller than $\xi^{(+)}/\xi^{(-)}$ by the factor $\pi^2\omega^2/4$, and the term with $\xi^{(-)}/\xi^{(+)}$ is beyond LL~\cite{Kirschner:1983di}. For this reason, Eq.~\eqref{eq:KLp} has no octet term. A color-octet gluon pair therefore contributes at LL only if it couples to $q-\bar q$ and has positive signature.

\begin{figure*}[t]
\centering
\includegraphics[page=6]{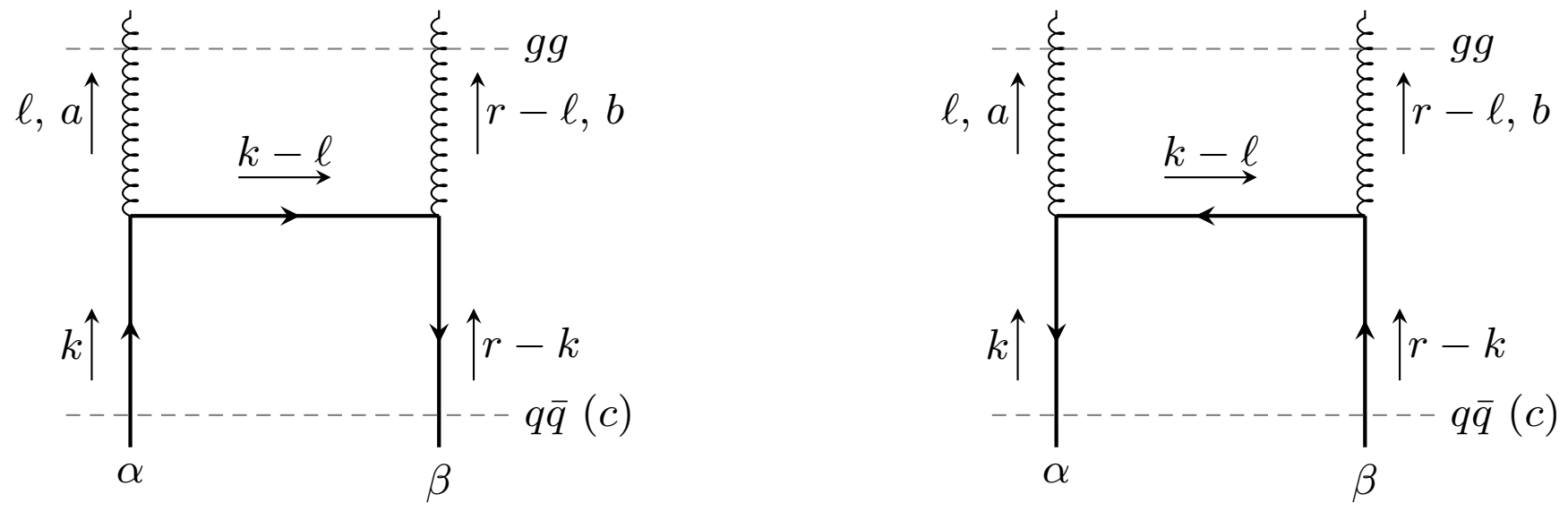}
\caption{The rung between a $t$-channel $q\bar q$ octet with color $c$ and a gluon pair, for a parent quark (a) and a parent antiquark (b). The momentum $r$ flows through the $t$ channel. The labels $\alpha$ and $\beta$ are the colors of the two fermion lines.}
\label{fig:rung}
\end{figure*}

Only $(gg)_{8_s}$ satisfies these conditions. A $q\bar q$ pair emits a gluon pair at the rung of Fig.~\ref{fig:rung}. For a color-octet $q\bar q$ pair, the color tensor of the rung is ${\rm tr}(T^bT^aT^c)$ for a parent quark [Fig.~\ref{fig:rung}(a)] and $-{\rm tr}(T^aT^bT^c)$ for a parent antiquark [Fig.~\ref{fig:rung}(b)]. By Eqs.~\eqref{eq:tr3} and \eqref{eq:tr3rev}, the color tensors of the parent combinations are
\begin{align}
&{\rm tr}(T^bT^aT^c)+{\rm tr}(T^aT^bT^c)=\frac12\,d^{abc}\qquad(q-\bar q),\label{eq:rungdiff}\\
&{\rm tr}(T^bT^aT^c)-{\rm tr}(T^aT^bT^c)=-\frac i2\,f^{abc}\qquad(q+\bar q).\label{eq:rungsum}
\end{align}
For a color-singlet $q\bar q$ pair, the color tensor of the rung is ${\rm tr}(T^bT^a)$ for a parent quark and ${\rm tr}(T^aT^b)$ for a parent antiquark. Both tensors equal $\delta^{ab}/2$ and cancel in $q-\bar q$. The state $(gg)_{8_s}$ therefore couples only to $q-\bar q$, and $(gg)_{8_a}$ and $(gg)_1$ couple only to $q+\bar q$. Color combinations of the two gluons that are symmetric under $a\leftrightarrow b$ have positive signature (Appendix~\ref{app:signature}), so that $(gg)_{8_s}$ has positive signature. Because gluons couple to every flavor with the same strength, $(gg)_1$ contributes only to the sea part of $q+\bar q$ and not to the valence part $\Pns^{+}$ (Sec.~\ref{sec:valsea}).

Because the outermost $t$-channel intermediate pair is a color singlet, the color octet $(gg)_{8_s}$ appears only between rungs inside the ladder. We consider that the conventional IREE miss $(gg)_{8_s}$ for this reason.

\subsection{The \KLG{}}
\label{sec:eqs}
The \KLG{} contain $(gg)_{8_s}$ among the color-octet $t$-channel intermediate pairs. We denote by $Q_j$ a color-octet $t$-channel intermediate pair $q_j\bar q_j$. In the \KLG{}, the positive-signature octet partial wave is a matrix $\Amat_8^+$ on the states $Q_1,\dots,Q_{n_f},(gg)_{8_s}$, and the negative-signature singlet partial wave is the matrix $\Amat_1^-$ of Eq.~\eqref{eq:readout_s}. The \KLG{} read
\begin{align}
\omega\Amat_1^-(\omega)&=2C_Fa_s\,\mathbf 1-\frac{8C_Fa_s}{\omega}\,(\Amat_8^+)_{QQ}(\omega)+[\Amat_1^-(\omega)]^2,\label{eq:KLGm}\\
\omega\Amat_8^+(\omega)&=a_s\Mmat_8+2C_Aa_s\,\frac{d\Amat_8^+(\omega)}{d\omega}+[\Amat_8^+(\omega)]^2.\label{eq:KLG8}
\end{align}
The matrix $(\Amat_8^+)_{QQ}$ is the block of $\Amat_8^+$ on $Q_1,\dots,Q_{n_f}$, and $\mathbf 1$ is the $n_f\times n_f$ unit matrix. The elements of the Born matrix $\Mmat_8$ on $Q_1,\dots,Q_{n_f},(gg)_{8_s}$ satisfy (Sec.~\ref{sec:coef})
\begin{align}
M_{Q_iQ_k}&=(2C_F-C_A)\,\delta_{ik},\label{eq:BornQQ}\\
M_{(gg)_{8_s}(gg)_{8_s}}&=-2C_A,\label{eq:BornGG}\\
M_{Q_i(gg)_{8_s}}M_{(gg)_{8_s}Q_k}&=-\frac{2(N_c^2-4)}{N_c}.\label{eq:BornQG}
\end{align}
The off-diagonal elements $M_{Q_i(gg)_{8_s}}$ and $M_{(gg)_{8_s}Q_k}$ depend on the normalization of the color tensor $d^{abc}$ of $(gg)_{8_s}$. Multiplying this tensor by a nonzero constant $\lambda$ multiplies one of them by $\lambda$ and the other by $1/\lambda$. Only the block $(\Amat_8^+)_{QQ}$ of $\Amat_8^+$ enters Eq.~\eqref{eq:KLGm}. Every contribution to this block enters and leaves $(gg)_{8_s}$ the same number of times. The off-diagonal elements of $\Mmat_8$ therefore enter $\Amat_1^-$ only through the product~\eqref{eq:BornQG}, and this product does not depend on the normalization of $d^{abc}$. If $M_{Q_i(gg)_{8_s}}$ and $M_{(gg)_{8_s}Q_k}$ are set to zero, Eqs.~\eqref{eq:KLGm} and \eqref{eq:KLG8} reduce to the conventional IREE~\eqref{eq:KLm} and \eqref{eq:KL8} for every flavor.

\begin{figure*}[t]
\centering
\includegraphics[page=4]{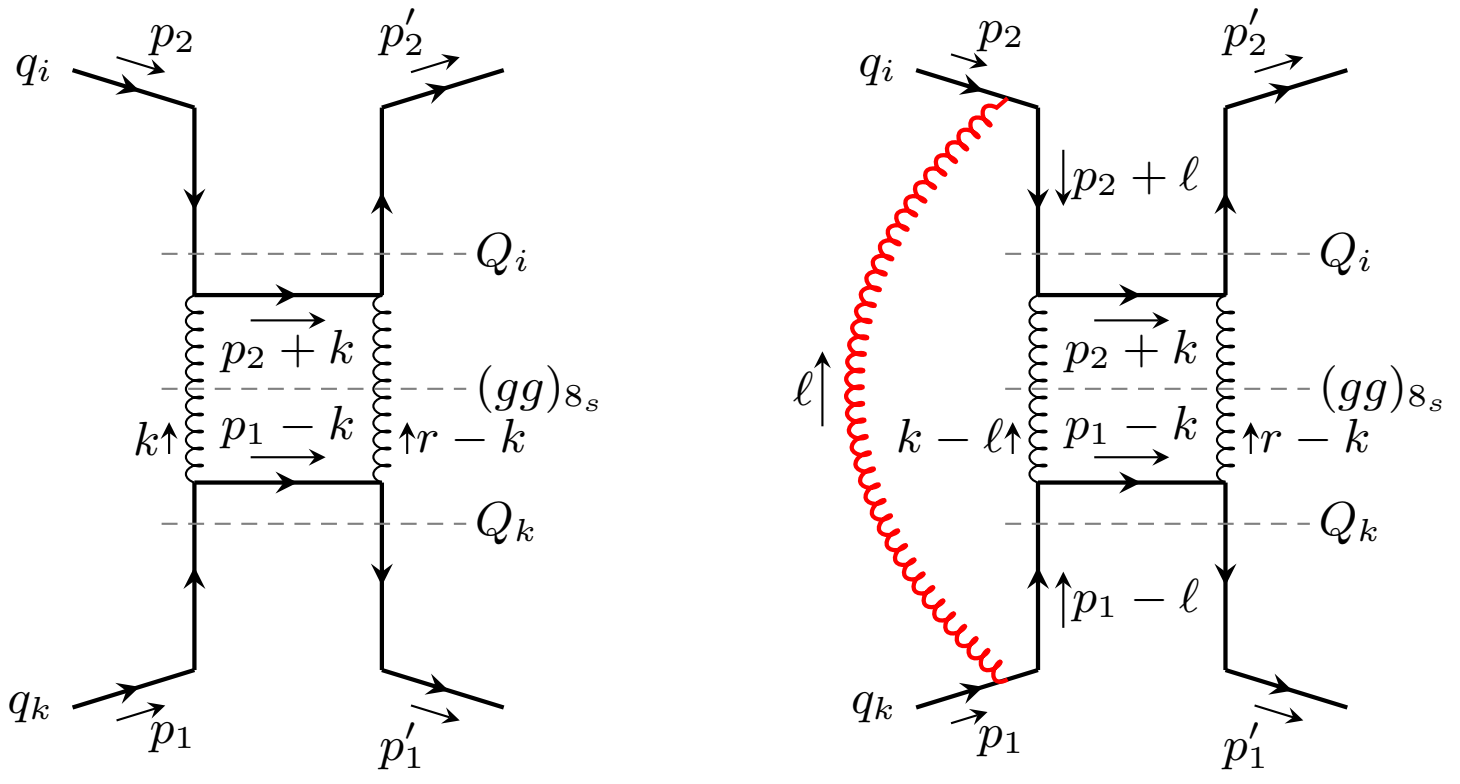}
\caption{Contributions of the path $Q_k\to(gg)_{8_s}\to Q_i$ to $\mathcal A_8^{(1){\rm sea}}$ (a) and $\mathcal A_1^{(2){\rm sea}}$ (b) [Eqs.~\eqref{eq:A8s2} and \eqref{eq:A1s3}], with $r=p_1-p_1'$. Dashed and red lines are as in Fig.~\ref{fig:terms}.}
\label{fig:path}
\end{figure*}

The path $Q_k\to(gg)_{8_s}\to Q_i$ produces the sea part of $\Amat_1^-$ (Fig.~\ref{fig:path}). In this path, the two gluons of $(gg)_{8_s}$ connect the fermion line of $q_k$ to the fermion line of $q_i$ [Fig.~\ref{fig:path}(a)]. For the same reason as for $(gg)_1$ (Sec.~\ref{sec:state}), the path takes the same value for all $i$ and $k$. The signature-changing term of Eq.~\eqref{eq:KLGm} feeds the path into $\Amat_1^-$ [Fig.~\ref{fig:path}(b)].

The Sudakov term has the coefficient $2C_Aa_s$ of Eq.~\eqref{eq:KL8} for every element of $\Amat_8^+$, and the signature-changing term has the coefficient $-8C_Fa_s/\omega$ of Eq.~\eqref{eq:KLm} for every element of $(\Amat_8^+)_{QQ}$. Both coefficients depend only on the color representation of the $t$ channel~\cite{Kirschner:1983di}, and $(gg)_{8_s}$ is a color octet like $Q_j$. The coefficient $2C_Aa_s$ of the Sudakov term is proportional to the value $C_A$ of the quadratic Casimir operator on a color octet. The color tensor of the external quarks in the path $Q_k\to(gg)_{8_s}\to Q_i$ is proportional to the color tensor $2\sum_aT^a\otimes T^a$ of the forward $qq\to qq$ amplitude with a color-octet $t$-channel intermediate pair:
\begin{equation}
\bigl[Q_k\to (gg)_{8_s}\to Q_i\bigr]_{\rm color}=\frac{N_c^2-4}{8N_c}\,2\sum_aT^a\otimes T^a.
\label{eq:pathPi8}
\end{equation}
In each product $T^a\otimes T^a$, the first factor is a matrix in the colors of the external quarks with the momenta $p_1$ and $p_1'$ in Fig.~\ref{fig:path}, and the second factor is a matrix in the colors of the external quarks with the momenta $p_2$ and $p_2'$.

\subsection{Born coefficients}
\label{sec:coef}
\begin{figure*}[t]
\centering
\includegraphics[page=5]{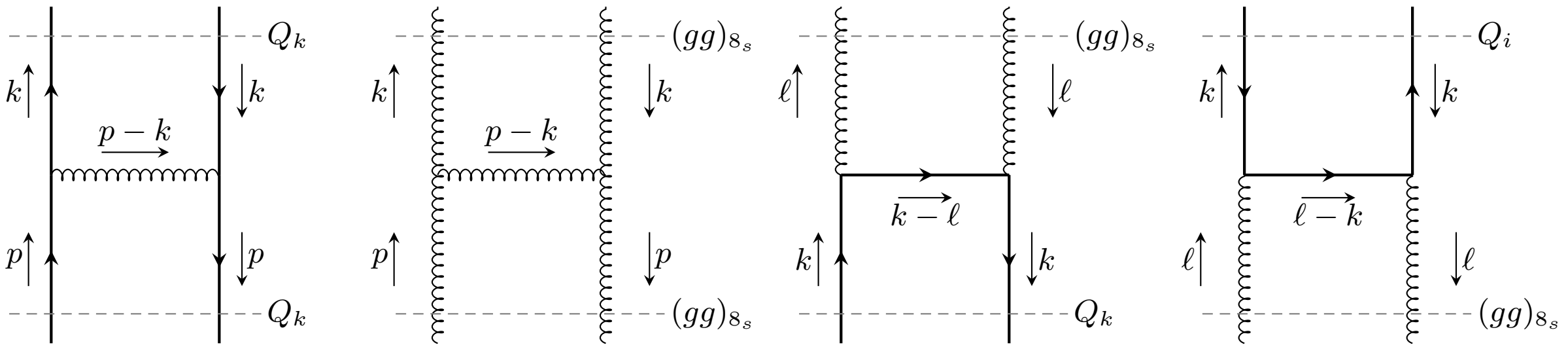}
\caption{Diagrams with one rung that give the Born coefficients~\eqref{eq:BornQQ}--\eqref{eq:BornQG}. Dashed lines are as in Fig.~\ref{fig:terms}.}
\label{fig:born}
\end{figure*}

We define the Born coefficient of a transition between two $t$-channel intermediate pairs as the coefficient of $x^0$ in the contribution of the diagram with one rung (Fig.~\ref{fig:born}) to the one-loop splitting function of this transition. Here $x$ is the fraction of the longitudinal momentum of the lower pair carried by the upper pair. The diagrams of Figs.~\ref{fig:born}(b) and (c) also give terms proportional to $x^{-1}$. The Mellin transform~\eqref{eq:mellin} turns $x^0$ into $1/N$ and $x^{-1}$ into $1/(N-1)$, which is regular at $N=0$. The terms proportional to $x^{-1}$ therefore do not contribute at LL and do not enter the Born coefficients. At $\omega=N$, the Born terms of Eqs.~\eqref{eq:KLGm} and \eqref{eq:KLG8} divided by $\omega$ are the Mellin transforms of $a_sx^0$ times the Born coefficients.

Because $(gg)_{8_s}$ is summed over the helicities of the two gluons, its Born coefficients follow from the unpolarized one-loop splitting functions. For color-singlet $t$-channel intermediate pairs, the diagrams of Fig.~\ref{fig:born} give the unpolarized one-loop splitting functions~\cite{Altarelli:1977zs}. For color-octet $t$-channel intermediate pairs, the calculation is the same except for the color factors (Appendix~\ref{app:colorG}). The results are Eqs.~\eqref{eq:BornQQ}--\eqref{eq:BornQG}.

\section{All-order solution}
\label{sec:closed}

The \KLG{} reduce to equations for the valence and sea parts of the partial waves and for the blocks that involve $(gg)_{8_s}$. The equations for the valence parts are the same as in the conventional IREE. We write the block $(\Amat_8^+)_{QQ}$ as
\begin{equation}
(\Amat_8^+)_{Q_iQ_k}(\omega)=\delta_{ik}\mathcal A_8^+(\omega)+\frac{\mathcal A_8^{\rm sea}(\omega)}{n_f}.
\label{eq:blocksQQ}
\end{equation}
The elements of $\Amat_8^+$ and $\Mmat_8$ between $Q_j$ and $(gg)_{8_s}$ do not depend on $j$, and we omit $j$ from their indices. Substituting Eqs.~\eqref{eq:readout_s} and \eqref{eq:blocksQQ} into Eqs.~\eqref{eq:KLGm} and \eqref{eq:KLG8} gives the reduced equations (Appendix~\ref{app:orders}).

\subsection{Closed form}
\label{sec:closedform}
The \KLG{} give $\mathcal A_1^{\rm sea}$ in closed form at fixed coupling. By Eqs.~\eqref{eq:Pv}, \eqref{eq:readout} and \eqref{eq:readout_s2}, the LL of $\cPns^{\rm val}(N)$ sum to $\mathcal A_1^{\rm val}(N)$, where
\begin{equation}
\mathcal A_1^{\rm val}(\omega)\equiv\mathcal A_1^-(\omega)+\mathcal A_1^{\rm sea}(\omega).
\label{eq:A1v}
\end{equation}
We obtain $\mathcal A_1^{\rm sea}$ as the closed form of $\mathcal A_1^{\rm val}$ minus the closed form of $\mathcal A_1^-$~\cite{Blumlein:1996dd}.

Multiplying Eq.~\eqref{eq:KLGm} from the right by the column vector $(1,\dots,1)^{\rm T}$ gives
\begin{equation}
\omega\mathcal A_1^{\rm val}(\omega)=2C_Fa_s-\frac{8C_Fa_s}{\omega}\,\mathcal A_8^{\rm val}(\omega)+[\mathcal A_1^{\rm val}(\omega)]^2,
\label{eq:A1veq}
\end{equation}
where $\mathcal A_8^{\rm val}(\omega)=\mathcal A_8^+(\omega)+\mathcal A_8^{\rm sea}(\omega)$. The root of Eq.~\eqref{eq:A1veq} that vanishes as $a_s\to0$ is
\begin{equation}
\mathcal A_1^{\rm val}(\omega)=\frac{\omega-\Delta^{\rm val}(\omega)^{1/2}}{2},
\label{eq:A1vclosed}
\end{equation}
where $\Delta^{\rm val}$ is the discriminant of Eq.~\eqref{eq:A1veq},
\begin{equation}
\Delta^{\rm val}(\omega)=\omega^2-8C_Fa_s\left(1-\frac{4\,\mathcal A_8^{\rm val}(\omega)}{\omega}\right).
\label{eq:Fv}
\end{equation}

We define the flavor-symmetric combination $Q_{\rm sym}=n_f^{-1/2}\sum_jQ_j$ of the color-octet $t$-channel intermediate pairs. By Eq.~\eqref{eq:blocksQQ}, the element $(\Amat_8^+)_{Q_{\rm sym}Q_{\rm sym}}$ is $\mathcal A_8^{\rm val}$. The Born matrix $\Mmat_8$ maps $Q_{\rm sym}$ and $(gg)_{8_s}$ onto linear combinations of $Q_{\rm sym}$ and $(gg)_{8_s}$, and so does every other term of Eq.~\eqref{eq:KLG8}. On these two states, Eq.~\eqref{eq:KLG8} becomes the Riccati equation
\begin{equation}
\begin{split}
&2C_Aa_s\,\frac{d\Amat_8^{\rm val}(\omega)}{d\omega}\\
&\quad=\omega\Amat_8^{\rm val}(\omega)-a_s\Mmat_8^{\rm val}-\bigl[\Amat_8^{\rm val}(\omega)\bigr]^2
\end{split}
\label{eq:riccati2}
\end{equation}
for the $2\times2$ block $\Amat_8^{\rm val}$ of $\Amat_8^+$ on $Q_{\rm sym}$ and $(gg)_{8_s}$ with the Born matrix
\begin{equation}
\Mmat_8^{\rm val}=\begin{pmatrix}2C_F-C_A&\sqrt{n_f}\,M_{Q(gg)_{8_s}}\\\sqrt{n_f}\,M_{(gg)_{8_s}Q}&-2C_A\end{pmatrix}.
\label{eq:M2}
\end{equation}
The solution with $\Amat_8^{\rm val}(\omega)\to a_s\Mmat_8^{\rm val}/\omega$ at large $\omega$ is (Appendix~\ref{app:riccati})
\begin{equation}
\begin{split}
&\Amat_8^{\rm val}(\omega)=\omega\,\mathbf 1_2\\
&\quad-\sqrt{2C_Aa_s}\;\mathbf V\,{\rm diag}\!\left(\frac{D_{\nu_1+1}(z)}{D_{\nu_1}(z)},\frac{D_{\nu_2+1}(z)}{D_{\nu_2}(z)}\right)\mathbf V^{-1},
\end{split}
\label{eq:A8closed}
\end{equation}
where $\mathbf 1_2$ is the $2\times2$ unit matrix, $z=\omega/\sqrt{2C_Aa_s}$, the constant matrix $\mathbf V$ diagonalizes $\Mmat_8^{\rm val}$ as $\mathbf V^{-1}\Mmat_8^{\rm val}\mathbf V=2C_A\,{\rm diag}(\nu_1,\nu_2)$, and $D_\nu(z)$ is the parabolic cylinder function~\cite{Olver:2010ouy}. In SU(3), $\nu_1$ and $\nu_2$ are
\begin{equation}
\nu_{1,2}=\frac{-19\pm i\sqrt{120\,n_f-289}}{36}.
\label{eq:nu12}
\end{equation}
If $\Mmat_8^{\rm val}$ is replaced by the number $2C_F-C_A$, Eq.~\eqref{eq:A8closed} becomes the solution $\mathcal A_8^+$ of Eq.~\eqref{eq:KL8}~\cite{Bartels:1995iu,Blumlein:1996dd}.

Equation~\eqref{eq:KLm} gives
\begin{equation}
\mathcal A_1^-(\omega)=\frac{\omega-\Delta^-(\omega)^{1/2}}{2},
\label{eq:A1mclosed}
\end{equation}
where $\Delta^-$ is the discriminant of Eq.~\eqref{eq:KLm}~\cite{Blumlein:1996dd}. Equation~\eqref{eq:A1v} then gives
\begin{equation}
\mathcal A_1^{\rm sea}(\omega)=\frac{\Delta^-(\omega)^{1/2}-\Delta^{\rm val}(\omega)^{1/2}}{2},
\label{eq:A1sclosed}
\end{equation}
where $\mathcal A_8^{\rm val}$ in $\Delta^{\rm val}$ is the element $(\Amat_8^{\rm val})_{Q_{\rm sym}Q_{\rm sym}}$ of Eq.~\eqref{eq:A8closed}. The expansion of Eq.~\eqref{eq:A1sclosed} in powers of $a_s/\omega^2$ has zero radius of convergence. For large $n$, the ratio of the coefficient of $a_s^{n+1}/\omega^{2n+1}$ to that of $a_s^n/\omega^{2n-1}$ behaves as $-4C_An$. The LL of $\cPns^{\rm sea}(N)$ therefore require the closed form~\eqref{eq:A1sclosed}, not a truncation of this expansion.

\subsection{Intercept}
\label{sec:intercept}
Because $D_\nu(z)$ is an entire function of $z$~\cite{Olver:2010ouy}, Eq.~\eqref{eq:A8closed} is singular only at the zeros of $D_{\nu_1}(z)$ and $D_{\nu_2}(z)$. Besides these zeros, the singularities of $\mathcal A_1^{\rm val}(\omega)$ with ${\rm Re}\,\omega>0$ are the branch points of Eq.~\eqref{eq:A1vclosed} at the zeros of $\Delta^{\rm val}(\omega)$. The position of the rightmost singularity of an LL anomalous dimension is called its intercept. We write $\omega_0^{\rm val}$ for the intercept of $\mathcal A_1^{\rm val}(\omega)$. The intercept gives the small-$x$ behavior $x^{-\omega_0^{\rm val}}$ of the LL of $\Pns^{\rm val}(x)$, up to powers of $\ln(1/x)$. Because $\mathcal A_8^{\rm val}(\omega)/\omega$ and $\Delta^{\rm val}(\omega)/(2C_Aa_s)$ depend on $\omega$ and $a_s$ only through $z$ [Eqs.~\eqref{eq:A8closed} and \eqref{eq:Fv}], the ratio $\omega_0^{\rm val}/\sqrt{2C_Aa_s}$ depends only on $N_c$ and $n_f$. In SU(3) with $n_f=3$, $4$ and $5$, the intercept is the largest real zero of $\Delta^{\rm val}(\omega)$, and Table~\ref{tab:intercept} gives $\omega_0^{\rm val}/\sqrt{2C_Aa_s}$.
In the conventional IREE, $\omega_0^{\rm val}$ is the intercept $\omega_0^-$ of $\mathcal A_1^-$ [Eq.~\eqref{eq:KLzero}]. The value of $\omega_0^-$ in Table~\ref{tab:intercept} agrees with that in Ref.~\cite{Bartels:1995iu}. In the \KLG{}, $\omega_0^{\rm val}$ is larger than $\omega_0^-$ for $n_f=3$, $4$ and $5$ (Table~\ref{tab:intercept}), and the LL of $\Pns^{\rm val}(x)$ therefore grow faster at small $x$ than in the conventional IREE. In Eq.~\eqref{eq:A1sclosed}, the rightmost singularities of the terms $\Delta^-(\omega)^{1/2}$ and $\Delta^{\rm val}(\omega)^{1/2}$ are at $\omega=\omega_0^-$ and $\omega=\omega_0^{\rm val}$, respectively. Because $\omega_0^-<\omega_0^{\rm val}$, the rightmost singularity of $\mathcal A_1^{\rm sea}(\omega)$ is at $\omega=\omega_0^{\rm val}$. The LL of $\Pns^{\rm sea}(x)$ therefore grow as $x^{-\omega_0^{\rm val}}$ up to powers of $\ln(1/x)$, faster than those of $\Pns^-(x)$. Accordingly, the ratio of the all-order LL of $\Pns^{\rm sea}(x)$ to those of $\Pns^-(x)$ increases as $x$ decreases [Fig.~\ref{fig:resumx}(b)].
\begin{table}
\caption{The intercept $\omega_0^{\rm val}$ in SU(3) in units of $\sqrt{2C_Aa_s}$. In the conventional IREE, $\omega_0^{\rm val}$ is the intercept $\omega_0^-$ of $\mathcal A_1^-$ and does not depend on $n_f$.}
\label{tab:intercept}
\begin{ruledtabular}
\begin{tabular}{ccccc}
&conventional IREE&\multicolumn{3}{c}{\KLG{}}\\
\cline{3-5}
$n_f$&&3&4&5\\
\hline
$\omega_0^{\rm val}/\sqrt{2C_Aa_s}$&1.389&1.426&1.436&1.447
\end{tabular}
\end{ruledtabular}
\end{table}

\subsection{\texorpdfstring{$\Pns^{\rm sea}$}{P\_ns\textasciicircum sea} in \texorpdfstring{$x$}{x} space}
\label{sec:xspace}
By Eq.~\eqref{eq:readout_s2}, the sum of the LL of $\Pns^{\rm sea}(x)$ at all orders is the inverse Mellin transform of Eq.~\eqref{eq:A1sclosed} at $\omega=N$. This transform turns each term $a_s^n/N^{2n-1}$ of the expansion of Eq.~\eqref{eq:A1sclosed} into $a_s^n\ln^{2n-2}x/(2n-2)!$. Because the coefficients grow as stated after Eq.~\eqref{eq:A1sclosed}, the factor $1/(2n-2)!$ makes the sum of the transformed terms converge.

\begin{figure}[t]
\centering
\includegraphics[width=\columnwidth]{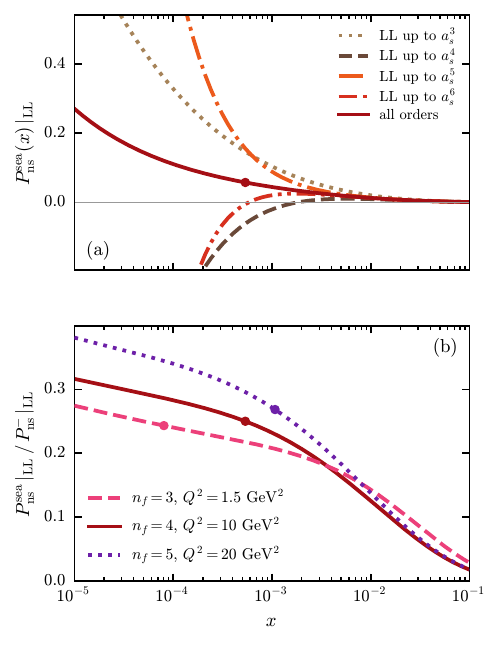}
\caption{(a) The LL of $\Pns^{\rm sea}(x)$ at all orders and up to $a_s^3$, $a_s^4$, $a_s^5$ and $a_s^6$ for $n_f=4$ and $Q^2=10$\,GeV$^2$. (b) Ratio of the all-order LL of $\Pns^{\rm sea}(x)$ to that of $\Pns^-(x)$ for $n_f=3$, $4$ and $5$ at $Q^2=1.5$, $10$ and $20$\,GeV$^2$, respectively. Dots mark $x=x_{\rm EIC}$.}
\label{fig:resumx}
\end{figure}

For the numerical values, we use $n_f=4$ and $Q^2=10$\,GeV$^2$. The EIC predictions of Refs.~\cite{Adamiak:2021ppq,Adamiak:2023yhz} are given at $Q^2=10$\,GeV$^2$. We write $x_{\rm EIC}=Q^2/(0.95\,s_{ep})$ for the lower limit of $x$ in the EIC studies of Ref.~\cite{AbdulKhalek:2021gbh}, where $\sqrt{s_{ep}}=140$\,GeV is the highest electron--proton center-of-mass energy of the EIC. For $Q^2=10$\,GeV$^2$, $x_{\rm EIC}$ is $5.4\times10^{-4}$. We use $x_{\rm EIC}$ as a reference value of small $x$. The values at $x=x_{\rm EIC}$ given below are LL results at fixed coupling, and those for $s-\bar s$ and $F_3$ are for a quark target. These values indicate the size of the contribution of $(gg)_{8_s}$ and are not predictions for measurements at the EIC. We obtain $\alpha_s(Q^2)$ from $\alpha_s(M_Z^2)=0.1180$ at the $Z$-boson mass $M_Z=91.188$\,GeV~\cite{ParticleDataGroup:2024cfk} by four-loop running in the modified minimal subtraction ($\overline{\rm MS}$) scheme, with $n_f=3$ below the charm-quark mass $m_c=1.273$\,GeV and $n_f=5$ above the bottom-quark mass $m_b=4.183$\,GeV~\cite{ParticleDataGroup:2024cfk}, and without matching corrections. This running gives $\alpha_s(10\,{\rm GeV}^2)=0.248$. In Fig.~\ref{fig:resumx}(b), we also use $n_f=3$ with $Q^2=1.5$\,GeV$^2<m_c^2$ and $n_f=5$ with $Q^2=20$\,GeV$^2>m_b^2$. At these two values of $Q^2$, $\alpha_s(Q^2)$ is $0.398$ and $0.220$, and $x_{\rm EIC}$ is $8.1\times10^{-5}$ and $1.1\times10^{-3}$, respectively.

Figure~\ref{fig:resumx}(a) shows the $x$ dependence of the all-order LL of $\Pns^{\rm sea}(x)$ and of its truncations at $a_s^3$ to $a_s^6$ (Sec.~\ref{sec:fixed}). Figure~\ref{fig:resumx}(b) shows, for $n_f=3$, $4$ and $5$, the ratio of the all-order LL
\begin{equation}
\frac{\Pns^{\rm sea}}{\Pns^-}=\frac{\Pns^{\rm val}|_{\rm ext}-\Pns^{\rm val}|_{\rm conv}}{\Pns^{\rm val}|_{\rm conv}},
\label{eq:ratio}
\end{equation}
where ext and conv denote the extended and the conventional IREE. The equality follows from Eqs.~\eqref{eq:A1v} and \eqref{eq:KLzero}, because the LL of $\Pns^-$ are the same in the two IREE (Sec.~\ref{sec:closed}).

\subsection{Strange-quark asymmetry}
\label{sec:ssbar}
We solve the DGLAP equation at fixed coupling with the all-order LL of the splitting functions given by the \KLG{} (Sec.~\ref{sec:closedform}). The solution gives the asymmetry $s-\bar s$ that the evolution generates from $s=\bar s$ at the initial scale. We write $q_{\rm ns}^{\rm val}$ for the total valence distribution. By Eq.~\eqref{eq:readout_s}, the moments of the valence distributions $q_i-\bar q_i$ evolve at LL as
\begin{equation}
\begin{split}
\frac{\partial(q_i-\bar q_i)(N,Q^2)}{\partial\ln Q^2}={}&\mathcal A_1^-(N)\,(q_i-\bar q_i)(N,Q^2)\\
&+\frac{\mathcal A_1^{\rm sea}(N)}{n_f}\,q_{\rm ns}^{\rm val}(N,Q^2).
\end{split}
\label{eq:valevol}
\end{equation}
The last term of Eq.~\eqref{eq:valevol} is the same for every flavor $i$. By Eq.~\eqref{eq:A1v}, the sum of Eq.~\eqref{eq:valevol} over $i$ shows that $q_{\rm ns}^{\rm val}$ evolves with $\mathcal A_1^{\rm val}(N)$. Subtracting $1/n_f$ times this sum from Eq.~\eqref{eq:valevol} shows that $(q_i-\bar q_i)-q_{\rm ns}^{\rm val}/n_f$ evolves with $\mathcal A_1^-(N)$. The solution of Eq.~\eqref{eq:valevol} from $Q^2=\mu^2$ is therefore
\begin{equation}
\begin{split}
&(q_i-\bar q_i)(N,Q^2)=\left(\frac{Q^2}{\mu^2}\right)^{\mathcal A_1^{\rm val}(N)}\frac{q_{\rm ns}^{\rm val}(N,\mu^2)}{n_f}\\
&\quad+\left(\frac{Q^2}{\mu^2}\right)^{\mathcal A_1^-(N)}\left[(q_i-\bar q_i)(N,\mu^2)-\frac{q_{\rm ns}^{\rm val}(N,\mu^2)}{n_f}\right].
\end{split}
\label{eq:valsol}
\end{equation}
If $s=\bar s$ at $Q^2=\mu^2$, Eq.~\eqref{eq:valsol} gives
\begin{equation}
(s-\bar s)(N,Q^2)=\left[1-\left(\frac{Q^2}{\mu^2}\right)^{-\mathcal A_1^{\rm sea}(N)}\right]\frac{q_{\rm ns}^{\rm val}(N,Q^2)}{n_f}.
\label{eq:ssbar}
\end{equation}
In Eq.~\eqref{eq:ssbar}, $q_{\rm ns}^{\rm val}(N,Q^2)$ is the solution
\begin{equation}
q_{\rm ns}^{\rm val}(N,Q^2)=\left(\frac{Q^2}{\mu^2}\right)^{\mathcal A_1^{\rm val}(N)}q_{\rm ns}^{\rm val}(N,\mu^2)
\label{eq:qvalsol}
\end{equation}
of the evolution of the total valence distribution.

By Eqs.~\eqref{eq:dataN} and \eqref{eq:readout_s2}, the first nonzero term of Eq.~\eqref{eq:ssbar} is the LL $8Da_s^3N^{-5}\ln(Q^2/\mu^2)\,q_{\rm ns}^{\rm val}(N,\mu^2)$ of the three-loop asymmetry of Ref.~\cite{Catani:2004nc}. In the conventional IREE, Eqs.~\eqref{eq:KLzero} and \eqref{eq:ssbar} give $s-\bar s=0$ at every power of $a_s$.

The factor in square brackets in Eq.~\eqref{eq:ssbar} does not depend on the input distributions. For simplicity, we take as the target a quark of a flavor other than $s$. At $Q^2=\mu^2$, the distributions in this target quark satisfy $s=\bar s=0$ and $q_{\rm ns}^{\rm val}(x,\mu^2)=\delta(1-x)$. For any other target, $s-\bar s$ in $x$ space is the convolution of $q_{\rm ns}^{\rm val}(x,\mu^2)$ of that target with $s-\bar s$ for the target quark.

\begin{figure}[t]
\centering
\includegraphics[width=\columnwidth]{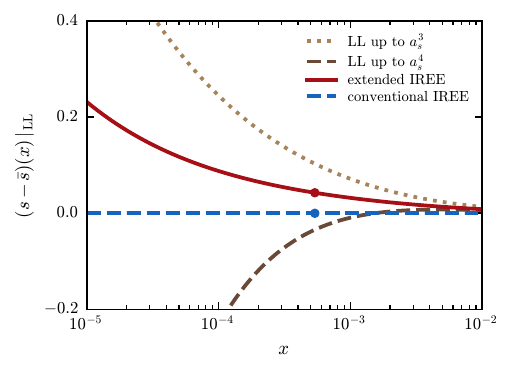}
\caption{The LL of $(s-\bar s)(x)$ at $Q^2=10$\,GeV$^2$ generated from $s=\bar s$ at $\mu^2=1$\,GeV$^2$ for a target quark of a flavor other than $s$ [Eq.~\eqref{eq:ssbar}]. The curves are the all-order results of the extended and the conventional IREE and the results with the LL of the splitting functions up to $a_s^3$ and up to $a_s^4$. Dots mark $x=x_{\rm EIC}$.}
\label{fig:ssbar}
\end{figure}

Figure~\ref{fig:ssbar} shows $s-\bar s$ for this target quark in the extended and the conventional IREE for $Q^2=10$\,GeV$^2$, $\mu^2=1$\,GeV$^2$ and the fixed coupling $a_s=\alpha_s(10\,{\rm GeV}^2)/(4\pi)$. Figure~\ref{fig:ssbar} also shows $s-\bar s$ from Eq.~\eqref{eq:ssbar} with $\mathcal A_1^-$ and $\mathcal A_1^{\rm sea}$ of the \KLG{} truncated at $a_s^3$ and at $a_s^4$. These truncations are the LL of $\cPns^-(N)$ and $\cPns^{\rm sea}(N)$ up to three loops and up to four loops (Sec.~\ref{sec:repro}).

At $x=x_{\rm EIC}$, the ratio of $s-\bar s$ to $q_{\rm ns}^{\rm val}/n_f$ in the \KLG{} is $0.19$, and the results with the truncations at $a_s^3$ and at $a_s^4$ are $2.4$ and $-0.80$ times the all-order LL, respectively (Fig.~\ref{fig:ssbar}). The results with the truncations at $a_s^3$ and at $a_s^4$ have opposite signs at $x=x_{\rm EIC}$. Therefore, neither the conventional IREE nor the fixed-order splitting functions known at present determine the LL of $s-\bar s$ at $x=x_{\rm EIC}$.

\subsection{Structure function \texorpdfstring{$F_3$}{F\_3}}
\label{sec:f3}
By the method of Ref.~\cite{Bartels:1995iu}, we obtain the all-order LL of the moments of $F_3$ summed over neutrino and antineutrino scattering on a quark target. We write $\mathcal F_{3,k}(N,Q^2/\mu^2)$ for the moments~\eqref{eq:FN} of $F_3$ for a target quark of flavor $k$. In Fig.~\ref{fig:readout} with a target quark of flavor $k$, the upper part is the amplitude for a target quark of the flavor $i$ of the softest $t$-channel intermediate pair, and the lower part is the element $(\Amat_1^-)_{ik}$. Equation~\eqref{eq:ireeF} therefore becomes
\begin{equation}
-\frac{\partial\mathcal F_{3,k}(N,Q^2/\mu^2)}{\partial\ln\mu^2}=\sum_i\mathcal F_{3,i}(N,Q^2/\mu^2)\,(\Amat_1^-)_{ik}(N).
\label{eq:ireeF3}
\end{equation}

\begin{figure}[t]
\centering
\includegraphics[page=7]{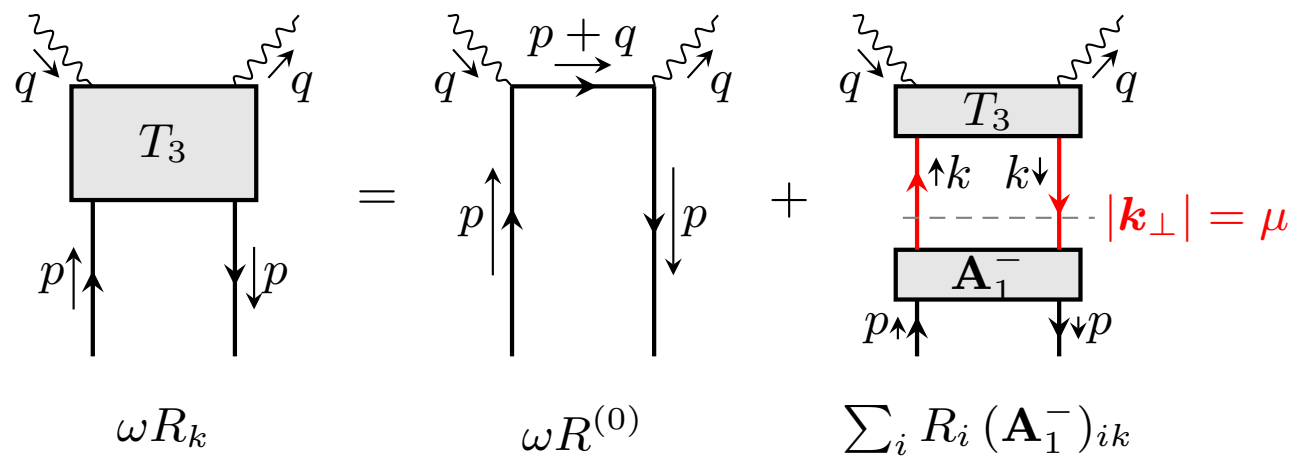}
\caption{Diagrams of the three terms of Eq.~\eqref{eq:phiF3}, with each term below its diagram. Red lines are the softest $t$-channel intermediate pair.}
\label{fig:phiF3}
\end{figure}

We write the partial wave of Eq.~\eqref{eq:TR} for a target quark of flavor $k$ as $R_k(\omega,Q^2/\mu^2)$. Equation~\eqref{eq:ireeF3} gives $\mathcal F_{3,k}$ from its value at $Q^2=\mu^2$, and we obtain this value from $R_k(\omega,1)$ as follows. At $Q^2=\mu^2$, the amplitude depends on $\mu$ only through $\rho=\ln(s/\mu^2)$, and the derivative $-\partial/\partial\ln\mu^2$ at fixed $s$ therefore multiplies the partial wave of each diagram by $\omega$ (Sec.~\ref{sec:pw}). We write $R^{(0)}(\omega,1)$ for the partial wave of the Born diagram. Because $F_3$ is summed over neutrino and antineutrino scattering, the Born diagram is the same for every flavor $k$ of the target quark, and its contribution to the derivative is $\omega R^{(0)}(\omega,1)$. Every other diagram is divided by its softest $t$-channel intermediate pair as in Fig.~\ref{fig:readout} and contributes the product of the partial waves of the two parts. Hence
\begin{equation}
\omega R_k(\omega,1)=\omega R^{(0)}(\omega,1)+\sum_iR_i(\omega,1)\,(\Amat_1^-)_{ik}(\omega).
\label{eq:phiF3}
\end{equation}
Figure~\ref{fig:phiF3} shows the diagrams of the three terms of Eq.~\eqref{eq:phiF3}.

By Eqs.~\eqref{eq:readout_s} and \eqref{eq:A1v}, the sum of the elements $(\Amat_1^-)_{ik}$ over $i$ does not depend on $k$:
\begin{equation}
\sum_i(\Amat_1^-)_{ik}(\omega)=\mathcal A_1^{\rm val}(\omega).
\label{eq:colsum}
\end{equation}
Because $R^{(0)}$ does not depend on $k$, Eq.~\eqref{eq:colsum} implies that the solution of Eq.~\eqref{eq:phiF3} at each power of $a_s$ does not depend on $k$ either:
\begin{equation}
R_k(\omega,1)=\frac{\omega R^{(0)}(\omega,1)}{\omega-\mathcal A_1^{\rm val}(\omega)}.
\label{eq:Rk}
\end{equation}
By Eq.~\eqref{eq:colsum}, Eq.~\eqref{eq:ireeF3} gives the factor $(Q^2/\mu^2)^{\mathcal A_1^{\rm val}(N)}$ for every $k$. By Eq.~\eqref{eq:FR}, the moments are $-\sin(\pi N)\,R_k(N,1)/(4\pi)$ times this factor. The moments therefore do not depend on $k$, and we drop the index $k$ below. At LL, $N$ is of the order of $\sqrt{a_s}$, and we approximate $\sin(\pi N)$ by $\pi N$. In the normalization of Eq.~\eqref{eq:Tdecomp}, the Born term of these moments is $1$~\cite{Moch:2008fj}, and the Born partial wave is therefore $R^{(0)}(N,1)=-4/N$. We then obtain
\begin{equation}
\mathcal F_3(N,Q^2/\mu^2)=\frac{N}{N-\mathcal A_1^{\rm val}(N)}\left(\frac{Q^2}{\mu^2}\right)^{\mathcal A_1^{\rm val}(N)}.
\label{eq:F3}
\end{equation}
In the conventional IREE, $\mathcal A_1^{\rm val}$ in Eq.~\eqref{eq:F3} is replaced by $\mathcal A_1^-$ [Eq.~\eqref{eq:KLzero}].

Equation~\eqref{eq:F3} is the result in the factorization scheme defined by the cutoff $\mu$. At $Q^2=\mu^2$, the factor $N/[N-\mathcal A_1^{\rm val}(N)]$ is the coefficient function in this scheme. The LL of this coefficient function converted to the $\overline{\rm MS}$ scheme agree with those of the fixed-order $\overline{\rm MS}$ coefficient function up to three loops, including the $d^{abc}$ term~\cite{Moch:2008fj} (Appendix~\ref{app:msbar}). The conversion does not change $F_3$ of Eq.~\eqref{eq:F3}, because the conversion multiplies the coefficient function by a factor and divides the total valence distribution of the target by the same factor.

By Eq.~\eqref{eq:A1vclosed}, $N-\mathcal A_1^{\rm val}(N)$ is $[N+\Delta^{\rm val}(N)^{1/2}]/2$. Because $\Delta^{\rm val}(N)^{1/2}$ has a non-negative real part, $N-\mathcal A_1^{\rm val}(N)$ does not vanish for ${\rm Re}\,N>0$, and the rightmost singularity of Eq.~\eqref{eq:F3} is the branch point $N=\omega_0^{\rm val}$. The term proportional to $(N-\omega_0^{\rm val})^{1/2}$ in the expansion of Eq.~\eqref{eq:F3} about $N=\omega_0^{\rm val}$ gives, for $\ln(1/x)\to\infty$,
\begin{equation}
\begin{split}
F_3(x,Q^2/\mu^2)\simeq{}&\frac{[\Delta^{\rm val\prime}(\omega_0^{\rm val})]^{1/2}}{2\sqrt\pi}\left(\frac{2}{\omega_0^{\rm val}}+\ln\frac{Q^2}{\mu^2}\right)\\
&\times\left(\frac{Q^2}{\mu^2}\right)^{\omega_0^{\rm val}/2}\!\frac{x^{-\omega_0^{\rm val}}}{\ln^{3/2}(1/x)},
\end{split}
\label{eq:F3asym}
\end{equation}
where the prime denotes $d/d\omega$.

\begin{figure}[t]
\centering
\includegraphics[width=\columnwidth]{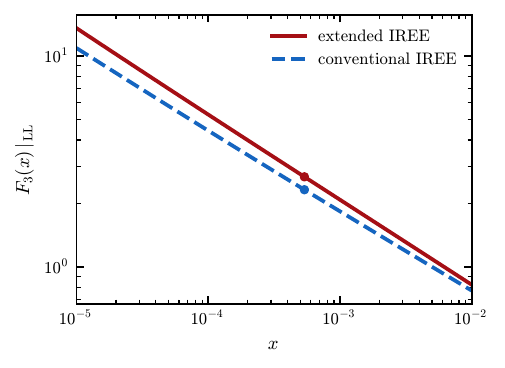}
\caption{The all-order LL of $F_3(x,Q^2/\mu^2)$ of a quark target [Eq.~\eqref{eq:F3}] in the extended and the conventional IREE for $Q^2=10$\,GeV$^2$ and $\mu^2=1$\,GeV$^2$. Dots mark $x=x_{\rm EIC}$.}
\label{fig:f3x}
\end{figure}

Figure~\ref{fig:f3x} shows the inverse Mellin transform $F_3(x,Q^2/\mu^2)$ of Eq.~\eqref{eq:F3} for $Q^2=10$\,GeV$^2$, $\mu^2=1$\,GeV$^2$ and the fixed coupling $a_s=\alpha_s(10\,{\rm GeV}^2)/(4\pi)$. At $x=x_{\rm EIC}$, $F_3$ in the \KLG{} is $1.15$ times that in the conventional IREE.

\section{Fixed orders: tests and predictions}
\label{sec:fixed}

We obtain the coefficients of the expansion of Eq.~\eqref{eq:A1sclosed} in powers of $a_s$ by solving the reduced equations (Appendix~\ref{app:orders}) for each power of $a_s$. Collecting the terms without $\delta_{ik}$ after substituting Eqs.~\eqref{eq:readout_s} and \eqref{eq:blocksQQ} into Eqs.~\eqref{eq:KLGm} and \eqref{eq:KLG8} gives the sea parts
\begin{align}
\omega\mathcal A_8^{\rm sea}(\omega)={}&2C_Aa_s\,\frac{d\mathcal A_8^{\rm sea}(\omega)}{d\omega}+2\mathcal A_8^+(\omega)\mathcal A_8^{\rm sea}(\omega)\nonumber\\
&+[\mathcal A_8^{\rm sea}(\omega)]^2+n_f(\Amat_8^+)_{Q(gg)_{8_s}}(\Amat_8^+)_{(gg)_{8_s}Q},\label{eq:A8s}\\
\omega\mathcal A_1^{\rm sea}(\omega)={}&-\frac{8C_Fa_s}{\omega}\,\mathcal A_8^{\rm sea}(\omega)+2\mathcal A_1^-(\omega)\mathcal A_1^{\rm sea}(\omega)\nonumber\\
&+[\mathcal A_1^{\rm sea}(\omega)]^2.\label{eq:A1s}
\end{align}

\subsection{Tests at three and four loops}
\label{sec:repro}

The \KLG{} reproduce the three-loop coefficient in Eq.~\eqref{eq:dataN} for all $N_c$. The color factor $D$ is not an input of the \KLG{}. The function $\mathcal A_8^{\rm sea}$ starts at order $a_s^2$. At this order, the last term of Eq.~\eqref{eq:A8s} gives [Fig.~\ref{fig:path}(a)]
\begin{equation}
\mathcal A_8^{(1){\rm sea}}(\omega)=\frac{n_fM_{Q(gg)_{8_s}}M_{(gg)_{8_s}Q}}{\omega^3}.
\label{eq:A8s2}
\end{equation}
At order $a_s^3$, the signature-changing term of Eq.~\eqref{eq:A1s} gives [Fig.~\ref{fig:path}(b)]
\begin{equation}
\mathcal A_1^{(2){\rm sea}}(\omega)=-\frac{8C_F}{\omega^2}\,\mathcal A_8^{(1){\rm sea}}(\omega)=\frac{8\,n_fD}{\omega^5}.
\label{eq:A1s3}
\end{equation}

The \KLG{} also reproduce the four-loop coefficient in Eq.~\eqref{eq:dataN} for all $N_c$ and the absence of an $n_f^2D$ term in Eq.~\eqref{eq:data4}. At order $a_s^4$, the Sudakov term and $M_{(gg)_{8_s}(gg)_{8_s}}$ contribute to $\mathcal A_1^{\rm sea}$ for the first time. At order $a_s^3$, $\mathcal A_8^{\rm sea}$ is (Appendix~\ref{app:orders})
\begin{equation}
\mathcal A_8^{(2){\rm sea}}(\omega)=\frac{n_fM_{Q(gg)_{8_s}}M_{(gg)_{8_s}Q}}{\omega^5}\left(8C_F-18C_A\right).
\label{eq:A8s3}
\end{equation}
Equation~\eqref{eq:A1s} with $\mathcal A_1^{(0)-}(\omega)=2C_F/\omega$ then gives
\begin{equation}
\begin{split}
&\mathcal A_1^{(3){\rm sea}}(\omega)\\
&\quad=-\frac{8C_F}{\omega^2}\,\mathcal A_8^{(2){\rm sea}}(\omega)+\frac{2\mathcal A_1^{(0)-}(\omega)\mathcal A_1^{(2){\rm sea}}(\omega)}{\omega}\\
&\quad=n_fD\,\frac{96C_F-144C_A}{\omega^7}.
\end{split}
\label{eq:A1s4}
\end{equation}
A term with $n_f^2$ requires two paths through $(gg)_{8_s}$, and such terms first appear in $\mathcal A_1^{\rm sea}$ at order $a_s^5$.

\subsection{Predictions at five and six loops}
\label{sec:pred56}

The \KLG{} contain no adjustable parameter. Their LL coefficients beyond four loops are therefore predictions. We obtain the five- and six-loop LL coefficients by continuing the solution of Appendix~\ref{app:orders} to $a_s^5$ and $a_s^6$. The results are
\begin{align}
a_s^5\,\mathcal A_1^{(4){\rm sea}}(\omega)&=n_fD\,\frac{a_s^5}{\omega^9}\Bigl[672C_F^2-2656C_FC_A\nonumber\\
&\qquad+2872C_A^2-80\,n_f\,\frac{N_c^2-4}{N_c}\Bigr]\nonumber\\
&\notebelow{{}=n_fD\,\frac{a_s^5}{\omega^9}\Bigl[672C_F^2-2656C_FC_A}{(\KLG{}),}\label{eq:pred5}\\
a_s^6\,\mathcal A_1^{(5){\rm sea}}(\omega)&=n_fD\,\frac{a_s^6}{\omega^{11}}\Bigl[2688C_F^3-30016C_F^2C_A\nonumber\\
&\qquad+75680C_FC_A^2-65872C_A^3\nonumber\\
&\qquad+n_f\,\frac{N_c^2-4}{N_c}\left(4544C_A-1536C_F\right)\Bigr]\nonumber\\
&\notebelow{{}=n_fD\,\frac{a_s^6}{\omega^{11}}\Bigl[2688C_F^3-30016C_F^2C_A}{(\KLG{}).}\label{eq:pred6}
\end{align}
Through the correspondence between $x$ space and Mellin space used for Eq.~\eqref{eq:dataN}, Eqs.~\eqref{eq:pred5} and \eqref{eq:pred6} at $\omega=N$ give the following small-$x$ terms of $\Pns^{\rm sea}$ in SU(3):
\begin{align}
&\Pns^{(4){\rm sea}}(x)\Big|_{\ln^8x}=n_f\left(\frac{6157}{3402}-\frac{25}{1701}\,n_f\right)\ln^8x\nonumber\\
&\notebelow{\Pns^{(4){\rm sea}}(x)\Big|_{\ln^8x}=n_f\left(\frac{6157}{3402}-\frac{25}{1701}\,n_f\right)\ln^8x}{(\KLG{}),}\label{eq:pred5x}\\
&\Pns^{(5){\rm sea}}(x)\Big|_{\ln^{10}x}=n_f\left(-\frac{115211}{91854}+\frac{362}{15309}\,n_f\right)\ln^{10}x\nonumber\\
&\notebelow{\Pns^{(5){\rm sea}}(x)\Big|_{\ln^{10}x}=n_f\left(-\frac{115211}{91854}+\frac{362}{15309}\,n_f\right)\ln^{10}x}{(\KLG{}).}\label{eq:pred6x}
\end{align}

In the \KLG{}, the Sudakov term of the element $(\Amat_8^+)_{(gg)_{8_s}(gg)_{8_s}}$ and the product of two such elements in the ladder term first contribute to $\Pns^{\rm sea}$ at five loops. We derive this Sudakov term and this product from the same attachments of the softest gluon as the corresponding gluon terms of the IREE for the polarized flavor singlet~\cite{Bartels:1996wc}. At large $N_c$, the LL gluon--gluon anomalous dimensions of the IREE for the polarized flavor singlet and of the small-$x$ helicity evolution of Ref.~\cite{Kovchegov:2015pbl} as revised in Ref.~\cite{Cougoulic:2022gbk} disagree from four loops on~\cite{Borden:2023ugd}. We therefore consider that a five-loop calculation of $\Pns^{\rm sea}$ can be one of the clues for judging this disagreement at fixed order.

\section{Discussion}
\label{sec:discussion}

We have added the symmetric color-octet two-gluon state $(gg)_{8_s}$ to the conventional IREE. The \KLG{} reproduce the three- and four-loop LL of $\Pns^{\rm sea}$ for all $N_c$ with no adjustable parameter. The intercept $\omega_0^{\rm val}$ of the \KLG{} is larger than the intercept $\omega_0^-$ of the conventional IREE for $n_f=3$, $4$ and $5$ (Table~\ref{tab:intercept}). The LL of $\Pns^{\rm val}(x)$ therefore grow faster at small $x$ than in the conventional IREE. To our knowledge, no other method in the literature gives the LL of $\Pns^{\rm sea}$ at all orders. We consider that the \KLG{} are the most promising approach to $\Pns^{\rm sea}$ at small $x$.

For a quark target at $x=x_{\rm EIC}$, $(gg)_{8_s}$ generates an asymmetry $s-\bar s$ equal to 19\% of $q_{\rm ns}^{\rm val}/n_f$ and increases the LL of $F_3$ by 15\% (Figs.~\ref{fig:ssbar} and \ref{fig:f3x}). We consider that $(gg)_{8_s}$ cannot be neglected in extrapolations of the total valence distribution and of $F_3$ to small $x$. The term with $\mathcal A_1^{\rm sea}$ in Eq.~\eqref{eq:valevol} is the same for every $q_i-\bar q_i$. With input distributions for $q_i-\bar q_i$ of all flavors, the \KLG{} therefore also give the all-order LL of $u-\bar u$ and $d-\bar d$.

At three to six loops and at large $N_c$ with fixed $n_f$, the ratio of the LL of $\Pns^{\rm sea}$ to those of $\Pns^-$ is proportional to $n_f/N_c$ [Eqs.~\eqref{eq:KLm}, \eqref{eq:KL8}, \eqref{eq:A1s3}, \eqref{eq:A1s4}, \eqref{eq:pred5} and \eqref{eq:pred6}]. The contribution of $(gg)_{8_s}$ therefore vanishes at large $N_c$ and is absent from small-$x$ evolution equations derived in the large-$N_c$ limit. In SU(3) with $n_f=4$, however, the three-loop LL of $\Pns^{\rm sea}$ are $4.1$ times those of $\Pns^-$ [Eqs.~\eqref{eq:KLm}, \eqref{eq:KL8} and \eqref{eq:A1s3}]. We therefore consider that the contribution of $(gg)_{8_s}$ cannot be neglected in QCD.

The results of Sec.~\ref{sec:closed} are limited to the LL at fixed coupling. The conventional IREE with the color-octet partial wave replaced by its Born term have been solved with running coupling~\cite{Ermolaev:2000sg}. For a proton target, $s-\bar s$ and $F_3$ are the convolutions of the input distributions with the results for a quark target. We leave the solution of the \KLG{} with running coupling and the evaluation of these convolutions for future work.

\begin{acknowledgments}
The author thanks Kenji Fukushima and Tomoya Uji for useful discussions.
The author used Anthropic's Claude Opus 5.5 to assist with the numerical code and the preparation of the manuscript, and checked all results.
This work was supported by the International Graduate Program of Innovation for Intelligent World (IIW), World-leading Innovative Graduate Study (WINGS) Program, The University of Tokyo.
\end{acknowledgments}

\appendix

\section{Color factors of the Born coefficients}
\label{app:colorG}

This appendix derives the Born coefficients~\eqref{eq:BornQQ}--\eqref{eq:BornQG}. For color-singlet and color-octet $t$-channel intermediate pairs, we compute the Born coefficient of each diagram of Fig.~\ref{fig:born} as the product of its color factor and its kinematic factor (Table~\ref{tab:born}). We obtain the kinematic factors from the one-loop splitting functions.
In the normalization of Eq.~\eqref{eq:conv}, the unpolarized one-loop splitting functions~\cite{Altarelli:1977zs} are, for $x<1$,
\begin{equation}
\begin{aligned}
P_{qq}^{(0)}(x)&=2C_F\,\frac{1+x^2}{1-x},\\
P_{gg}^{(0)}(x)&=4C_A\left(\frac{1}{1-x}+\frac{1}{x}-2+x-x^2\right),\\
P_{gq}^{(0)}(x)&=2C_F\left(\frac{2}{x}-2+x\right),\\
P_{qg}^{(0)}(x)&=2n_f\left[x^2+(1-x)^2\right].
\end{aligned}
\label{eq:LOx0}
\end{equation}
Their coefficients of $x^0$ are $2C_F$, $-4C_A$, $-4C_F$ and $2n_f$, respectively.
The factor $2n_f$ of $P_{qg}^{(0)}$ counts the quark and the antiquark of each flavor, and the coefficient of $x^0$ for one pair $q_k\bar q_k$ is $2$. For color-singlet $t$-channel intermediate pairs, the Born coefficients of the diagrams of Figs.~\ref{fig:born}(a)--(d) are the coefficients $2C_F$, $-4C_A$, $-4C_F$ and $2$ of $x^0$, respectively, and the kinematic factors in Table~\ref{tab:born} follow.

For Figs.~\ref{fig:born}(a) and (b), the color factors for color-octet $t$-channel intermediate pairs in Table~\ref{tab:born} follow from $\sum_cT^cT^eT^c=(C_F-C_A/2)\,T^e$ and $\sum_{a',b',c}f^{aa'c}f^{bb'c}d^{a'b'e}=(C_A/2)\,d^{abe}$. For Figs.~\ref{fig:born}(c) and (d), the product of the two color factors is the squared color amplitude of the rung of Fig.~\ref{fig:rung}, with the normalized color tensors $\sqrt2\,T^c$ of the $q\bar q$ pair and $\sqrt{N_c/(N_c^2-4)}\,d^{abc}$ of $(gg)_{8_s}$. With Eq.~\eqref{eq:tr3} and $\sum_{a,b}d^{abc'}d^{abc}=(N_c^2-4)\,\delta^{cc'}/N_c$, this product is
\begin{equation}
\biggl|\sqrt{\frac{2N_c}{N_c^2-4}}\,\sum_{a,b}d^{abc}\,{\rm tr}(T^bT^aT^c)\biggr|^2=\frac{N_c^2-4}{8N_c}
\label{eq:colamp8}
\end{equation}
for a fixed $c$. For color-singlet $t$-channel intermediate pairs, the normalized color tensors $\mathbf 1_{N_c}/\sqrt{N_c}$ of the $q\bar q$ pair and $\delta^{ab}/\sqrt{N_c^2-1}$ of $(gg)_1$ give $C_F/2$ instead, where $\mathbf 1_{N_c}$ is the $N_c\times N_c$ unit matrix.
\begin{table}[t]
\caption{Kinematic factors, color factors (color) and Born coefficients (Born) of the diagrams of Fig.~\ref{fig:born} for color-singlet and color-octet $t$-channel intermediate pairs. The Born coefficient is the product of the color factor and the kinematic factor. For color-octet pairs, the Born coefficients are those of Eqs.~\eqref{eq:BornQQ}--\eqref{eq:BornQG}.}
\label{tab:born}
\begin{ruledtabular}
\begin{tabular}{lccccc}
Fig.~\ref{fig:born}&kinematic&\multicolumn{2}{c}{color singlet}&\multicolumn{2}{c}{color octet}\\
&factor&color&Born&color&Born\\
\hline
(a)&$2$&$C_F$&$2C_F$&$C_F-C_A/2$&$2C_F-C_A$\\
(b)&$-4$&$C_A$&$-4C_A$&$C_A/2$&$-2C_A$\\
(c)$\times$(d)&$-16$&$C_F/2$&$-8C_F$&$\dfrac{N_c^2-4}{8N_c}$&$-\dfrac{2(N_c^2-4)}{N_c}$
\end{tabular}
\end{ruledtabular}
\end{table}

\section{Signature of two-gluon states}
\label{app:signature}
\begin{figure}[t]
\centering
\includegraphics[page=8,width=\columnwidth]{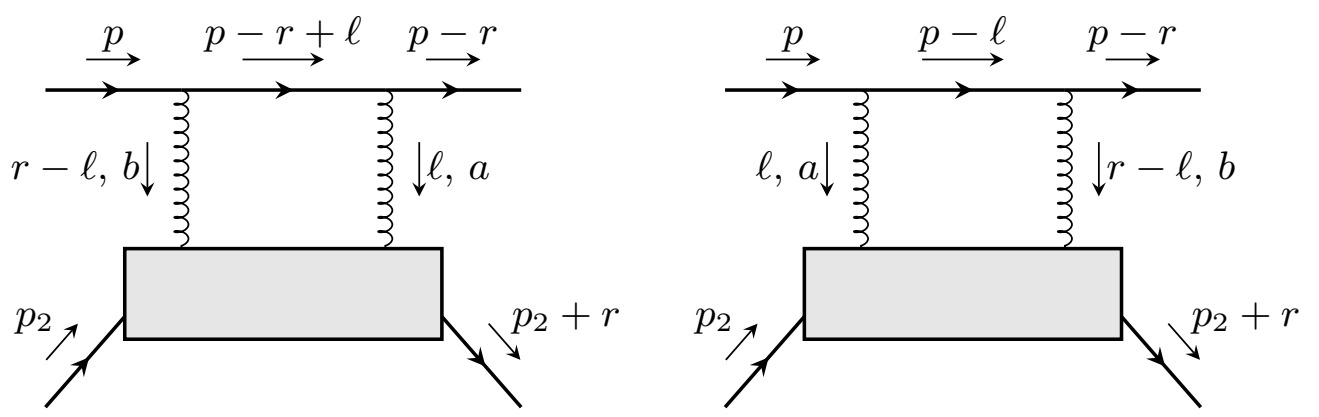}
\caption{The two orderings of two gluons with colors $a$ and $b$ on a quark line. The gray box is the part of the amplitude other than the quark line and the two gluons attached to it. Under each diagram are its color factor and its kinematic function.}
\label{fig:orderings}
\end{figure}

This appendix shows that a color combination of two gluons that is symmetric under $a\leftrightarrow b$ has positive signature. Figure~\ref{fig:orderings} shows the two orderings of two gluons with colors $a$ and $b$ on a quark line. We write $K(s)$ for the kinematic function of the ordering with the color factor $T^aT^b$ [Fig.~\ref{fig:orderings}(a)]. The crossing $s\leftrightarrow u$ turns the ordering in Fig.~\ref{fig:orderings}(a) into that in Fig.~\ref{fig:orderings}(b). Because $u=-s$ at leading power in $1/s$, the kinematic function of the ordering in Fig.~\ref{fig:orderings}(b) is $K(-s)$. The amplitude is therefore
\begin{equation}
A(s)=T^aT^b\,K(s)+T^bT^a\,K(-s).
\label{eq:sig}
\end{equation}
The signature parts are the parts of $A$ that are symmetric and antisymmetric under $s\leftrightarrow u$ at fixed colors:
\begin{equation}
\begin{split}
A^{\pm}(s)&=\frac{A(s)\pm A(-s)}{2}\\
&=\frac12\left(T^aT^b\pm T^bT^a\right)\left[K(s)\pm K(-s)\right].
\end{split}
\label{eq:sigpm}
\end{equation}
In SU($N_c$), the color matrices in Eq.~\eqref{eq:sigpm} are
\begin{align}
T^aT^b+T^bT^a&=\frac{\delta^{ab}}{N_c}+d^{abc}T^c,\label{eq:TTsym}\\
T^aT^b-T^bT^a&=if^{abc}T^c.\label{eq:TTanti}
\end{align}
By Eqs.~\eqref{eq:sigpm}--\eqref{eq:TTanti}, the color singlet $(gg)_1$ and the color octet $(gg)_{8_s}$ with the tensor $d^{abc}$ have positive signature, and the color octet $(gg)_{8_a}$ with the tensor $f^{abc}$ has negative signature.

\section{\texorpdfstring{Solution at each power of $a_s$}{Solution at each power of a\_s}}
\label{app:orders}
This appendix gives the equations that we solve for each power of $a_s$ to obtain the coefficients of Sec.~\ref{sec:fixed}. We write $J$ for the $n_f\times n_f$ matrix with $J_{ik}=1$ for all $i$ and $k$. Every flavor matrix in Eqs.~\eqref{eq:KLGm} and \eqref{eq:KLG8} is a linear combination of $\mathbf 1$ and $J$. We use $J^2=n_fJ$ and $(\Amat_8^+)_{Q_i(gg)_{8_s}}(\Amat_8^+)_{(gg)_{8_s}Q_k}=(\Amat_8^+)_{Q(gg)_{8_s}}(\Amat_8^+)_{(gg)_{8_s}Q}J_{ik}$. The blocks of Eq.~\eqref{eq:KLG8} then give Eqs.~\eqref{eq:KL8} and \eqref{eq:A8s} and the equations
\begin{align}
&\omega(\Amat_8^+)_{Q(gg)_{8_s}}=a_sM_{Q(gg)_{8_s}}+2C_Aa_s\,\frac{d(\Amat_8^+)_{Q(gg)_{8_s}}}{d\omega}\nonumber\\
&\quad+\bigl[\mathcal A_8^{\rm val}+(\Amat_8^+)_{(gg)_{8_s}(gg)_{8_s}}\bigr]\,(\Amat_8^+)_{Q(gg)_{8_s}},\label{eq:AQG}\\
&\omega(\Amat_8^+)_{(gg)_{8_s}(gg)_{8_s}}=a_sM_{(gg)_{8_s}(gg)_{8_s}}\nonumber\\
&\quad+2C_Aa_s\,\frac{d(\Amat_8^+)_{(gg)_{8_s}(gg)_{8_s}}}{d\omega}\nonumber\\
&\quad+n_f(\Amat_8^+)_{Q(gg)_{8_s}}(\Amat_8^+)_{(gg)_{8_s}Q}+[(\Amat_8^+)_{(gg)_{8_s}(gg)_{8_s}}]^2.\label{eq:AGG}
\end{align}
The equation for $(\Amat_8^+)_{(gg)_{8_s}Q}$ is Eq.~\eqref{eq:AQG} with every subscript $Q(gg)_{8_s}$ replaced by $(gg)_{8_s}Q$. The Born terms give the coefficients of $a_s$
\begin{align}
&\mathcal A_8^{(0)+}(\omega)=\frac{2C_F-C_A}{\omega},\label{eq:first8}\\
&(\Amat_8^{(0)+})_{Q(gg)_{8_s}}(\omega)=\frac{M_{Q(gg)_{8_s}}}{\omega},\label{eq:firstQGapp}\\
&(\Amat_8^{(0)+})_{(gg)_{8_s}(gg)_{8_s}}(\omega)=-\frac{2C_A}{\omega}.\label{eq:firstGG}
\end{align}
Solving Eqs.~\eqref{eq:KLm}, \eqref{eq:KL8}, \eqref{eq:A8s}, \eqref{eq:A1s}, \eqref{eq:AQG}, \eqref{eq:AGG} and the equation for $(\Amat_8^+)_{(gg)_{8_s}Q}$ for each power of $a_s$ from these coefficients gives Eqs.~\eqref{eq:A8s2}--\eqref{eq:A1s4}, \eqref{eq:pred5} and \eqref{eq:pred6}.

\section{Closed-form solution of the Riccati equation}
\label{app:riccati}
This appendix derives the solution~\eqref{eq:A8closed} of the Riccati equation~\eqref{eq:riccati2}. The substitution
\begin{equation}
\Amat_8^{\rm val}(\omega)=2C_Aa_s\,\boldsymbol\Psi'(\omega)\boldsymbol\Psi(\omega)^{-1}
\label{eq:subst}
\end{equation}
turns Eq.~\eqref{eq:riccati2} into the linear equation
\begin{equation}
2C_Aa_s\,\boldsymbol\Psi''(\omega)-\omega\boldsymbol\Psi'(\omega)+\frac{\Mmat_8^{\rm val}}{2C_A}\,\boldsymbol\Psi(\omega)=0
\label{eq:linear}
\end{equation}
for a $2\times2$ matrix $\boldsymbol\Psi(\omega)$. With the matrix $\mathbf V$ of Eq.~\eqref{eq:A8closed}, the matrix $\boldsymbol\Psi=\mathbf V\,{\rm diag}(\psi_1,\psi_2)\,\mathbf V^{-1}$ solves Eq.~\eqref{eq:linear} if each $\psi_i$ solves
\begin{equation}
\frac{d^2\psi_i}{dz^2}-z\,\frac{d\psi_i}{dz}+\nu_i\psi_i=0.
\label{eq:hermite}
\end{equation}
Equation~\eqref{eq:hermite} has the two independent solutions $e^{z^2/4}D_{\nu_i}(z)$ and $e^{z^2/4}D_{-\nu_i-1}(iz)$~\cite{Olver:2010ouy}. For $|z|\to\infty$ with $|\arg z|<3\pi/4$, the leading term of the asymptotic expansion of $D_\nu(z)$ is $z^\nu e^{-z^2/4}$~\cite{Olver:2010ouy}. At large real $\omega$, both $z$ and $iz$ satisfy this condition, and the two solutions behave as
\begin{align}
&e^{z^2/4}D_{\nu_i}(z)\to z^{\nu_i},\nonumber\\
&e^{z^2/4}D_{-\nu_i-1}(iz)\to(iz)^{-\nu_i-1}e^{z^2/2}.
\label{eq:psiasym}
\end{align}
With $dz/d\omega=1/\sqrt{2C_Aa_s}$, the first solution gives
\begin{equation}
2C_Aa_s\frac{\psi_i'(\omega)}{\psi_i(\omega)}\to\frac{2C_Aa_s\nu_i}{\omega},
\label{eq:psiratio1}
\end{equation}
and any solution that contains the second solution gives $2C_Aa_s\psi_i'/\psi_i\to\omega$. With $\boldsymbol\Psi=\mathbf V\,{\rm diag}(\psi_1,\psi_2)\,\mathbf V^{-1}$, Eq.~\eqref{eq:subst} gives
\begin{equation}
\Amat_8^{\rm val}(\omega)=\mathbf V\,{\rm diag}\Bigl(2C_Aa_s\frac{\psi_1'}{\psi_1},\,2C_Aa_s\frac{\psi_2'}{\psi_2}\Bigr)\mathbf V^{-1}.
\label{eq:A8diag}
\end{equation}
The diagonalization $\mathbf V^{-1}\Mmat_8^{\rm val}\mathbf V=2C_A\,{\rm diag}(\nu_1,\nu_2)$ in Sec.~\ref{sec:closedform} gives
\begin{equation}
\frac{a_s\Mmat_8^{\rm val}}{\omega}=\mathbf V\,{\rm diag}\Bigl(\frac{2C_Aa_s\nu_1}{\omega},\,\frac{2C_Aa_s\nu_2}{\omega}\Bigr)\mathbf V^{-1}.
\label{eq:M8diag}
\end{equation}
By Eqs.~\eqref{eq:A8diag} and \eqref{eq:M8diag}, the large-$\omega$ behavior $\Amat_8^{\rm val}(\omega)\to a_s\Mmat_8^{\rm val}/\omega$ imposed in Sec.~\ref{sec:closedform} requires Eq.~\eqref{eq:psiratio1} for $i=1$ and $2$. The solution is therefore $\psi_i=e^{z^2/4}D_{\nu_i}(z)$. The relation $dD_\nu/dz=zD_\nu/2-D_{\nu+1}$~\cite{Olver:2010ouy} gives
\begin{equation}
2C_Aa_s\frac{\psi_i'(\omega)}{\psi_i(\omega)}=\omega-\sqrt{2C_Aa_s}\,\frac{D_{\nu_i+1}(z)}{D_{\nu_i}(z)},
\label{eq:psiratio}
\end{equation}
and substituting Eq.~\eqref{eq:psiratio} into Eq.~\eqref{eq:A8diag} gives Eq.~\eqref{eq:A8closed}.

\section{\texorpdfstring{Coefficient function of $F_3$ in the $\overline{\rm MS}$ scheme}{Coefficient function of F\_3 in the MS-bar scheme}}
\label{app:msbar}
This appendix converts the coefficient function $N/[N-\mathcal A_1^{\rm val}(N)]$ of Eq.~\eqref{eq:F3} from the cutoff scheme to the $\overline{\rm MS}$ scheme at LL. We keep the cutoff $\mu$ and work in $4-2\epsilon$ dimensions with the mass scale $\mu_\epsilon$ of dimensional regularization. Each loop then has the factor $(\boldsymbol k_\perp^2/\mu_\epsilon^2)^{-\epsilon}$, and the softest parton with $|\boldsymbol k_\perp|=\mu$ has the coupling
\begin{equation}
\hat a_s=a_s\left(\frac{\mu^2}{\mu_\epsilon^2}\right)^{-\epsilon}.
\label{eq:ahat}
\end{equation}
In $4-2\epsilon$ dimensions, the $qq\to qq$ amplitude has the mass dimension $2\epsilon$. In place of Eq.~\eqref{eq:pw}, we write
\begin{equation}
A_{1,8}^{\tau}(\rho)=\mu^{2\epsilon}\int\frac{d\omega}{2\pi i}\,e^{\omega\rho}\,\xi^{(\tau)}(\omega)\,\mathcal A_{1,8}^{\tau}(\omega),
\label{eq:pweps}
\end{equation}
where the partial waves $\mathcal A_{1,8}^{\tau}$ are functions of $\omega$, $\epsilon$ and $\hat a_s$. The amplitude $T_3$ is dimensionless, and Eq.~\eqref{eq:TR} keeps its form. The derivative $-\partial/\partial\ln\mu^2$ at fixed $s$ and $a_s$ acts on $\mu^{2\epsilon}$, on $\rho=\ln(s/\mu^2)$ and on $\hat a_s$. Equation~\eqref{eq:ahat} gives $-\partial\hat a_s/\partial\ln\mu^2=\epsilon\hat a_s$. The derivative therefore multiplies the partial waves in Eq.~\eqref{eq:pweps} by the operator
\begin{equation}
\omega-\epsilon+\epsilon\,\hat a_s\frac{\partial}{\partial\hat a_s},
\label{eq:opA}
\end{equation}
and the partial wave $R$ in Eq.~\eqref{eq:TR} by $\omega+\epsilon\,\hat a_s\,\partial/\partial\hat a_s$.

The $\overline{\rm MS}$ coefficient function obtained below depends on the partial waves only up to first order in $\epsilon$. At LL, the terms of first order in $\epsilon$ have one more power of $1/\omega$ than the terms at $\epsilon=0$. We assume that the terms of first order in $\epsilon$ on the right-hand sides of Eqs.~\eqref{eq:KLGm}, \eqref{eq:KLG8}, \eqref{eq:ireeF3} and \eqref{eq:phiF3} arise only from the replacement of $a_s$ by $\hat a_s$. With this assumption, Eqs.~\eqref{eq:KLGm}, \eqref{eq:KLG8} and \eqref{eq:phiF3} become
\begin{align}
&\Bigl(\omega-\epsilon+\epsilon\,\hat a_s\frac{\partial}{\partial\hat a_s}\Bigr)\Amat_1^-(\omega)\nonumber\\
&\quad=2C_F\hat a_s\,\mathbf 1-\frac{8C_F\hat a_s}{\omega}\,(\Amat_8^+)_{QQ}(\omega)+[\Amat_1^-(\omega)]^2,\label{eq:KLGmeps}\\
&\Bigl(\omega-\epsilon+\epsilon\,\hat a_s\frac{\partial}{\partial\hat a_s}\Bigr)\Amat_8^+(\omega)\nonumber\\
&\quad=\hat a_s\Mmat_8+2C_A\hat a_s\,\frac{d\Amat_8^+(\omega)}{d\omega}+[\Amat_8^+(\omega)]^2,\label{eq:KLG8eps}\\
&\Bigl(\omega+\epsilon\,\hat a_s\frac{\partial}{\partial\hat a_s}\Bigr)R_k(\omega,1)\nonumber\\
&\quad=\omega R^{(0)}(\omega,1)+\sum_iR_i(\omega,1)\,(\Amat_1^-)_{ik}(\omega),\label{eq:phiF3eps}
\end{align}
and Eq.~\eqref{eq:ireeF3} keeps its form with $\Amat_1^-$ as a function of $\hat a_s$. We write $\mathcal A^{(n)}(\omega,\epsilon)$ for the coefficient of $\hat a_s^{n+1}$ in a partial wave $\mathcal A$. The operator~\eqref{eq:opA} multiplies $\hat a_s^{n+1}\mathcal A^{(n)}$ by $\omega+n\epsilon$. Solving Eqs.~\eqref{eq:KLGmeps} and \eqref{eq:KLG8eps} for each power of $\hat a_s$ as in Appendix~\ref{app:orders} gives, in place of Eq.~\eqref{eq:A1s3},
\begin{equation}
\mathcal A_1^{(2){\rm sea}}(\omega,\epsilon)=\frac{8\,n_fD}{\omega^5}-\frac{24\,n_fD\,\epsilon}{\omega^6}+O(\epsilon^2).
\label{eq:A1s3eps}
\end{equation}

We set $\mu_\epsilon=Q$, so that $\hat a_s=a_s(Q^2/\mu^2)^{\epsilon}$. We write $C_\mu(N,\epsilon)$ for the moments of $F_3$ at $Q^2=\mu^2$, and $C_\mu(N,0)$ is $N/[N-\mathcal A_1^{\rm val}(N)]$. Equation~\eqref{eq:ireeF3} with the derivative at fixed $a_s$ then gives
\begin{equation}
\begin{split}
&\mathcal F_3(N,Q^2/\mu^2)=C_\mu(N,\epsilon)\\
&\quad\times\exp\left[\sum_{n\ge0}a_s^{n+1}\mathcal A_1^{(n){\rm val}}(N,\epsilon)\,\frac{(Q^2/\mu^2)^{(n+1)\epsilon}-1}{(n+1)\,\epsilon}\right].
\end{split}
\label{eq:F3eps}
\end{equation}
For $\epsilon<0$, the limit $\mu\to0$ removes the cutoff, and $(Q^2/\mu^2)^{(n+1)\epsilon}$ vanishes. Following Ref.~\cite{Davies:2022ofz}, we write the result of this limit as the product of the $\overline{\rm MS}$ coefficient function and a factor that contains only poles in $\epsilon$. The resummation of Ref.~\cite{Davies:2022ofz} does not cover the $x^0$ terms of $F_3$ summed over neutrino and antineutrino scattering. At LL, the exponent of this factor is the sum of the terms $-a_s^{n+1}\cPns^{(n){\rm val}}(N)/[(n+1)\epsilon]$. The poles of Eq.~\eqref{eq:F3eps} have this form with $\cPns^{(n){\rm val}}(N)$ equal to $\mathcal A_1^{(n){\rm val}}(N,0)$, and the remainder at $\epsilon=0$ is the $\overline{\rm MS}$ coefficient function
\begin{equation}
\begin{split}
C_{\overline{\rm MS}}(N)={}&\frac{N}{N-\mathcal A_1^{\rm val}(N)}\\
&\times\exp\left[-\sum_{n\ge0}\frac{a_s^{n+1}}{n+1}\,\frac{\partial\mathcal A_1^{(n){\rm val}}}{\partial\epsilon}(N,0)\right].
\end{split}
\label{eq:CMS}
\end{equation}

Up to $a_s^3$, the cutoff scheme and Eq.~\eqref{eq:CMS} give
\begin{align}
&\frac{N}{N-\mathcal A_1^{\rm val}(N)}=1+\frac{2C_Fa_s}{N^2}+\left(8C_AC_F-8C_F^2\right)\frac{a_s^2}{N^4}\nonumber\\
&\quad+\left(-24C_A^2C_F+128C_AC_F^2-120C_F^3+8n_fD\right)\frac{a_s^3}{N^6},\label{eq:Ccut3}\\
&C_{\overline{\rm MS}}(N)=1+\frac{2C_Fa_s}{N^2}+\left(12C_AC_F-14C_F^2\right)\frac{a_s^2}{N^4}\nonumber\\
&\quad+\left(-48C_A^2C_F+232C_AC_F^2-212C_F^3+16n_fD\right)\frac{a_s^3}{N^6}.\label{eq:CMS3}
\end{align}
In Eq.~\eqref{eq:CMS3}, the term $16n_fD$ is the sum of $8n_fD$ from Eq.~\eqref{eq:Ccut3} and $8n_fD$ from the exponential of Eq.~\eqref{eq:CMS} with Eq.~\eqref{eq:A1s3eps}. Through the correspondence $\ln^{2n-1}x\leftrightarrow-(2n-1)!/N^{2n}$, Eq.~\eqref{eq:CMS3} agrees with the LL of the one-, two- and three-loop $\overline{\rm MS}$ coefficient functions [Eqs.~(3.2), (3.17) and (3.20) of Ref.~\cite{Moch:2008fj}]. For $N_c=3$, Ref.~\cite{Moch:2008fj} sets the symbol $d^{abc}d_{abc}/n_c$ of its Eq.~(3.20) to $5n_f/18$. Equation~\eqref{eq:D} gives $n_fD/16=5n_f/18$ for $N_c=3$, and we therefore replace this symbol by $n_fD/16$. We consider that the agreement of the terms without $D$ at two and three loops supports the assumption on the terms of first order in $\epsilon$.

\bibliography{IREE_refs}

\end{document}